\documentstyle[twocolumn,aps,epsf,psfig]{revtex}

\begin{document}
\title{Correlated--electron theory of  strongly anisotropic 
metamagnets}

\author{K. Held$^1$, M. Ulmke$^1$, N. Bl\"umer$^2$,
  and D.~Vollhardt$^1$ }

\address{ $^1$Theoretische Physik III, 
Universit\"at Augsburg, D--86135 Augsburg, Germany}
 
\address{$^2$Department of Physics, 
University of Illinois at Urbana--Champaign, Urbana, IL 61801--3080, USA} 

\date{ \today}

\maketitle

\begin{abstract}
The microscopic origin of metamagnetism and
metamagnetic transitions in strongly anisotropic antiferromagnets is
investigated within a quantum mechanical theory of 
correlated electrons.
To this end the
Hubbard model with staggered magnetization ${\bf m}_{st}$ along an easy axis 
$\bf e$ in
a magnetic field ${\bf H} \parallel {\bf e}$ is studied both 
analytically and numerically
within the dynamical mean field theory (DMFT). At intermediate couplings
the self--consistent DMFT equations, which become exact in the limit of
large coordination number, are solved by finite temperature Quantum
Monte Carlo techniques. The temperature and magnetic field dependence of
the homogeneous and staggered magnetization are calculated and the
magnetic phase diagram is constructed. At half filling the metamagnetic
transitions are found to change from first order at low temperatures to
second order near the N\'eel temperature, implying the existence of a
multicritical point. Doping with holes or electrons has a strong
effect: the system becomes metallic, the electronic compressibility
increases and the critical temperatures and fields decrease.  These
results are related to known properties of insulating metamagnets such
as FeBr$_2$, metallic metamagnets such as UPdGe, and the giant and
colossal magnetoresistance found in a number of magnetic bulk systems.
\vspace{12pt}
\end{abstract}

\section{Introduction}
\label{intro}
While ferromagnetism is known since antiquity,
 antiferromagnetism was only discovered in this century.
It is not widely known that the concept of antiferromagnetic order was 
proposed independently by  N\'{e}el\cite{Neel32} in 1932
and Landau\cite{Lan33} in 1933.
Both sought to explain the, at that time, puzzling low temperature behavior 
of the magnetic susceptibility of certain materials:
of metals such as Cr and Mn in the case of N\'{e}el, and of insulators with 
layered structure such as the
chlorides of Cr, bivalent Fe, Co and Ni in the case of Landau.
While N\'{e}el correctly suggested the existence of interpenetrating 
sublattices in Cr and Mn with opposite magnetization \cite{Neel32},
Landau equally correctly predicted the existence of
stacks of ferromagnetically
ordered layers whose magnetization alternates from layer to layer \cite{Lan33}.
In both cases the total spontaneous magnetization adds up to zero.
Assuming the interlayer coupling to be weak, Landau \cite{Lan33}
argued that a relatively small magnetic field would be sufficient to modify 
the mutual orientation of the moments in each layer.
This leads to deviations from the linear dependence of the total moment on the
field, i.e.~to an anomalous increase of the susceptibility, and finally 
-- at high fields -- to a saturation of the magnetization. Such a behavior was 
indeed observed by Becquerel and van den Handel in 1939 \cite{1}
in the mineral
mesitite (carbonate of Fe and Mg) at low temperatures. Not being aware
of Landau's or N\'{e}el's work they could not
explain their observation in terms of
ferro- and paramagnetism, and therefore 
suggested for it the name metamagnetism \cite{2}. 

Qualitatively similar, but  even more drastic magnetization effects
 were later observed in many other systems of which
FeCl$_2$ and Dy$_3$Al$_5$O$_{12}$ (DAG) are well-studied
prototypes  \cite{Str77}.
These materials are insulators where the valence electrons
are localized at the Fe and Dy ions, respectively.
The resulting local moments order {antiferromagnetically} and are 
constrained to lie along an easy axis ${\mathbf e}$,
implying a strong anisotropy such that a spin--flop transition  cannot occur.
Under the influence of a large magnetic field 
${\mathbf H} \parallel {\mathbf e}$ 
the staggered magnetization vanishes in a first or second order 
phase transition, the so--called metamagnetic transition.
Apart from the thoroughly investigated 
materials mentioned above,  there are also conducting systems
in that class, e.g.~Uranium--based mixed systems 
\cite{Str77,Sechovsky94}, SmMn$_2$Ge$_2$ \cite{Bra94}, 
and  TbRh$_{2-x}$Ir$_x$Si$_2$ \cite{Iva95}.

Today the term ``metamagnetic transition'' is used
in a much wider sense \cite{4,5}, namely whenever the 
homogeneous susceptibility
$\chi (H)$ has a maximum at some value $H_c$, 
with $m(H)$ being strongly enhanced for $H > H_c$.
Metamagnetism is then found to be a rather common
phenomenon which occurs also in 
spin--flop antiferromagnets (e.g.~the 
parent compound of high-T$_c$ superconductivity, La$_2$CuO$_4$ \cite{5}), 
strongly exchange-enhanced paramagnets
(e.g.~TiBe$_2$, YCo$_2$ \cite{4}), heavy fermion and intermediate valence
systems (e.g.~CeRu$_2$Si$_2$, UPt$_3$ \cite{5}).

In this paper we will be concerned only with
metamagnetism in strongly anisotropic antiferromagnets,
several of which
are known to have a very interesting $H-T$ phase diagram
($H$:  {\it internal} magnetic field, $T$: temperature).
In particular, in the insulating systems one often finds a tricritical point 
at which the first order phase transition becomes second order.
Theoretical investigations of tricritical points
began with the work of Landau, who
described multicritical behavior within his phenomenological
theory of phase transitions\cite{Lan37}.
Clearly, a genuine understanding of the origin of tricritical points 
etc.~requires investigations on a more microscopic level. In the case of  
strongly anisotropic metamagnets those investigations where sofar
restricted to the insulating systems, such as  FeBr$_2$.
They are usually based on the
Ising model with more than one interaction,
in a magnetic field, on a simple cubic lattice \cite{7},  e.g. 
\begin{equation}
H = J \sum_{NN} S_iS_j - J' \sum_{NNN} S_i S_j - H \sum_i S_i \; ,
\label{Gl1}\label{isi}
\end{equation}
where the summations extend over the nearest neighbors (NN) and
next nearest neighbors (NNN) of every site.
For $J, J' > 0$ one has  an antiferromagnetic (AF) coupling between the $Z$
nearest neighbor spins and a ferromagnetic 
coupling between the $Z'$ next nearest neighbors \cite{8a}.
 It was  pointed out by 
Kincaid and Cohen \cite{8}
that in mean--field theory a tricritical point \cite{Griffiths70} 
as in Fig.~1a exists only for
$R \equiv Z'J'/(ZJ) > 3/5$, while for $R < 3/5$ this point separates
into  a critical endpoint (CE) and a bicritical endpoint (BCE)
(see Fig.~1b).
The latter behavior, especially the finite angle between the 
two transition lines at CE and the pronounced maximum at the 
second order line,
 is qualitatively very similar to that observed in FeBr$_2$ \cite{9}.
However, the first order line between CE and BCE 
has so far not been observed -- neither experimentally,
nor even theoretically when evaluating (\ref{Gl1})
beyond mean--field theory \cite{10,Selke95,Selke96}. Most recently, by
measurement of the excess magnetization and anomalous susceptibility
loss \cite{Azevedo95}, 
the specific heat \cite{Katori96}, and  by neutron
scattering \cite{Katsumata97} a strip--shaped regime
of strong {\it non--critical}  fluctuations in the $H_a - T$
plane (where $H_a$ is the applied field) of FeBr$_2$ was reported
which is at
least reminiscent of the {\it critical} 
line CE $\leftrightarrow$ BCE.
This unusual behavior was then also found theoretically by 
Selke \cite{Selke96} who evaluated (\ref{Gl1}) and also a more detailed model
\cite{Pleimling97} 
for weak ferromagnetic coupling ($R = 0.4$) 
and large $Z$ using Monte Carlo techniques.
There does not yet exist a microscopic theory for the 
conducting systems such as the Uranium--based mixed systems
\cite{Str77,Sechovsky94}, because this requires a fully
quantum mechanical treatment of \em itinerant, \em correlated electrons.
First steps in this direction, where genuine correlation effects were,
however, neglected, are the semi--phenomenological theories of Wohlfarth
and Rhodes \cite{Wohlfarth62} 
for metamagnetic phase transitions in paramagnets,
and of Moriya and Usami \cite{Moriya77} for the coexistence of ferro- and
antiferromagnetism in
itinerant electron systems.
It is the purpose of this paper to examine the origin of metamagnetism
in strongly anisotropic antiferromagnets from a microscopic, 
quantum mechanical point of view.
To this end we will study the Hubbard model in the presence of a strong
anisotropy direction, with a magnetic field along this direction,
since it is the simplest microscopic model that describes insulating
\em and \em metallic, spin--localized \em and \em bandlike antiferromagnets
with easy axis in an external magnetic field.

In Sec.~\ref{easy} the underlying model, the Hubbard model with
easy axis, is introduced and its validity is discussed. Then the dynamical 
mean--field theory which is used to investigate the correlation problem all 
the way from weak to strong coupling, as well as the Quantum Monte Carlo 
techniques employed to solve the coupled self--consistency equations,
are discussed in Sec.~\ref{dmft}. The results obtained for a half filled 
band and for finite doping are presented in Sec.~\ref{reshf}
and Sec.~\ref{resbhf}, respectively.
A discussion of the results in Sec.~\ref{dis} closes the presentation.

\vspace{3cm}

\begin{figure}[t]
\hspace{-4.9cm} \psfig{file=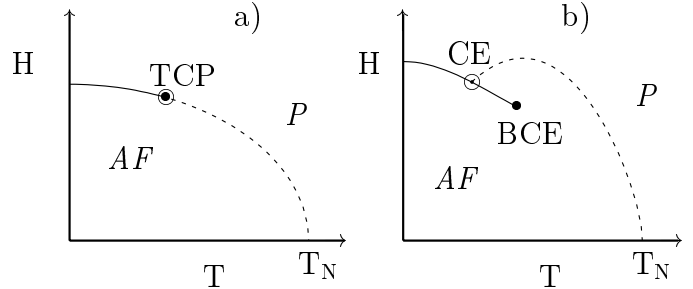,width=8.9cm}
\vspace{-9.5cm}\caption{
Schematic phase diagram, magnetic field $H$ vs.~temperature $T$,  for
a) a typical Ising-type metamagnet (TCP: tricritical point), 
b) the Ising model (\ref{isi}) in
mean--field theory with $R < 3/5$ (CE: critical endpoint, BCE:
bicritical endpoint). Solid lines:  first order transition, 
broken lines: second order
transition; AF: antiferromagnetic phase, P: paramagnetic phase.
\label{scheme}\\}
\end{figure}

\section{Hubbard Model with Easy Axis}
\label{easy}

The  Hubbard model \cite{Hubbard63} is the generic microscopic model
for itinerant and localized antiferromagnetism 
in correlated electron systems. 
For nearest neighbor hopping of electrons in the presence of a magnetic field 
it has the form
\begin{eqnarray}
\hat{H} =  & & -t \sum_{NN, \sigma} \hat{c}_{i \sigma}^+ \hat{c}_{j \sigma}
+ U \sum_i \hat{n}_{i \uparrow} \hat{n}_{i \downarrow} \nonumber \\
& &  -  \sum_{i \sigma} (\mu + \sigma H) \hat{n}_{i \sigma}  \; ,
\label{Gl2}\label{hub}
\end{eqnarray}
where operators carry a hat.
 From the exact, analytic solution in  dimensions $d = 1$ 
the (paramagnetic) ground state of this
model at half filling $(n=1)$ is known to exhibit metamagnetic behavior, 
i.e.~$\partial \chi/\partial H > 0$, up to saturation \cite{Takahashi69}.
However, 
this is entirely due to the convex shape of the density of states in $d = 1$
and occurs even at $U = 0$. 

In general the Hubbard model in the form of (\ref{hub}) cannot describe
metamagnetic behavior. The Hubbard model is isotropic in spin space and hence 
the antiferromagnetic phase described by it is isotropic, too.
Consequently, any finite magnetic field $\bf H$ will orient 
the staggered magnetization ${\bf m}_{st}$ {\em perpendicular} to itself. 
Real antiferromagnets are, however, never isotropic in spin space: the 
relativistic spin--orbit interaction $\bf \hat{L}  \cdot \hat{S}$, where 
 $\bf \hat{L}$ and  $\bf \hat{S}$ are operators for the orbital 
angular momentum and spin, respectively, transfers the anisotropy of 
{\em position} space (caused by the broken rotational symmetry of the 
crystal lattice) into {\em spin} space, producing one, or more, 
easy axes which
constrain the spins \cite{dipdip}. 
In an external magnetic field such a constraint leads to
metamagnetic transitions, with or without a spin--flop
depending on the strength of the spin--orbit interaction, 
as explained in Sec.~\ref{intro}.

A microscopic theory of strongly anisotropic antiferromagnets should 
ultimately be able to take into account the orbital degeneracy of the 
electrons and, by including the relativistic spin--orbit interaction
${\mathbf \hat L \cdot \hat S}$ in the Hamiltonian,
to generate an anisotropy axis within the model itself.
At present, this is technically not possible \cite{Kulakowski90}. 
Therefore we take the existence of the anisotropy axis ${\mathbf e}$
for granted: we employ the Hubbard model (\ref{hub}) 
and constrain the magnetic moments to lie along 
${\mathbf e} \parallel  {\mathbf H}$. 
By this approach the kinetic energy and the Coulomb interaction are treated 
microscopically, whereas the relativistic corrections are not.
We note that the relativistic corrections are of the order of $10^{-2}$ eV
and are thus small compared to energies of the order of $1$ eV
for kinetic and Coulomb energy.  
Therefore the existence of ${\mathbf e}$ and the correlation physics
described by the Hubbard model (\ref{hub}) are quite unrelated.
This justifies our approach where the existence of ${\mathbf e}$
is {\em a priori} assumed \cite{easyaxis}. 

\section{Dynamical mean--field theory ($D\to\infty$)}
\label{dmft}

For classical spin models (e.g.~the Ising model) 
it is well known that the Weiss
molecular field theory becomes exact in the limit of high spatial
dimensions ($d=\infty$). 
For lattice electrons this limit was introduced only recently
\cite{Met89}.
With the proper scaling of the hopping element in (\ref{Gl2}),
$t = t^*/\sqrt{Z}$ ($Z$ = number of nearest neighbors), 
it leads to  a quantum mechanical dynamical
mean--field theory (DMFT); for reviews see \cite{dv93,Georges96}.
The interacting lattice model then reduces to
a self--consistent single site 
problem of  electrons in an effective medium \cite{Jan91}, 
which may be described by a complex, frequency dependent
(i.e.~dynamical) self energy $\Sigma^\sigma(\omega)$ .
This problem is, in fact, equivalent to an Anderson impurity model 
complemented by a self--consistency condition \cite{Georges92,Jarrell92,Kot94}.

There are two limits in which the DMFT recovers well-known
{\em static} mean--field theories:
\begin{enumerate}

\item
{\em {Weak coupling:}}
In this situation the effective medium may be
approximated by a \em static \em field which is generated by the 
\em averaged \em densities of the electrons. 
This leads to the Hartree--Fock approximation, 
e.~g.~$\Sigma^\sigma(\omega)=U n_{-\sigma}$ in the homogeneous case, 
which is expected to give the 
qualitatively correct behavior at weak coupling.
The averaged densities $n_\sigma$ have to be 
determined self--consistently.

\item
{\em{Strong coupling at half filling:}}
Here model (\ref{Gl2}) can be mapped onto the antiferromagnetic 
spin 1/2 Heisenberg model for which the limit $d=\infty$ becomes
equivalent to the Weiss molecular field theory.
\end{enumerate}

The results obtained in  these two limits will be presented, and compared 
to the results for intermediate coupling, in Sec.~\ref{reshf}.

The DMFT has recently provided valuable 
insight into the physics of strongly correlated electron systems, e.g. the 
Mott--Hubbard transition \cite{Georges96,Jarrell92,Gebhard97} 
and transport properties \cite{20}. 
The effect of the magnetic field $H$ in (\ref{Gl2}) for $n = 1$ was 
also studied  \cite{21,22,23}. In particular, Laloux et al.~\cite{21} 
thoroughly investigated the magnetization behavior of the
paramagnetic phase, assuming the AF order to  be
suppressed. For $U = 3 \sqrt{2} t^*$  they find a first order
metamagnetic transition between the strongly correlated metal 
and the Mott insulator at a critical field $H \simeq 0.2 t^*$.
Giesekus and Brandt \cite{23}  took into
account the AF order. They considered the isotropic
case where the field  orients 
the staggered magnetization  perpendicular to itself, 
such that a metamagnetic transition cannot occur.

To investigate the metamagnetic phase transition from 
an antiferromagnet to a paramagnet we consider a bipartite ($A$-$B$) lattice
and allow for  symmetry breaking with respect to spin
$\sigma \in \{\uparrow, \downarrow\}=\{+,-\}$ and sublattice
$\alpha \in \{A,B\} = \{+,-\}$.
The  self energy $\Sigma^\sigma_{\alpha n}
\equiv\Sigma^\sigma_{\alpha}(i\omega_n)$,
with Matsubara frequencies $\omega_n=\pi T (2n+1)$, $n=0,\pm 1,\pm 2,\dots$
(using the convention $\hbar\equiv k_B \equiv 1$),
and the Green function $G^\sigma_{\alpha n}$ are determined
self--consistently by two sets of coupled equations \cite{Jan91,Georges92}:

\begin{eqnarray}
G_{\alpha n}^{\sigma} &= &
\int_{-\infty}^\infty d \epsilon \frac{N^0 (\epsilon)}
{z_{\alpha n}^{\sigma} - \epsilon^2/z_{-\alpha n}^\sigma}\label{dys}  \\
G_{\alpha n}^{\sigma} &= & 
- \langle \psi_{\sigma n}^{\phantom *} \psi_{\sigma n}^* 
\rangle_{A_\alpha} \; .\label{fi}
\label{Gl4}
\end{eqnarray}
Here $z_{\alpha n}^\sigma = i \omega_n +  \mu - \Sigma_{\alpha n}^\sigma$, 
and the thermal average of some operator 
${\cal O}[ \psi, \psi^*]$ in (\ref{Gl4}) is defined as a functional integral
over the Grassmann variables $\psi, \psi^*$, with 

\begin{equation}
\langle {\cal O} 
\rangle_{A_\alpha} = \frac{1}{Z_\alpha}
 \int {\cal D} [\psi ] {\cal D} [\psi^*] 
{\cal O[ \psi, \psi^*] }
e^{A_{\alpha}[ \psi, \psi^*,\Sigma, G] } ,
\end{equation} 
in terms of the single site action
\begin{eqnarray}
A_\alpha &=& \sum_{\sigma, n} \psi_{\sigma n}^* 
\big[ (G_{\alpha n}^\sigma)^{-1} + 
\Sigma_{\alpha n}^{\sigma} \big] \psi_{\sigma n} \nonumber \\
&&- U \int_0^\beta
d \tau \psi_\uparrow^* (\tau) \psi_\uparrow (\tau) 
 \psi_\downarrow^* (\tau) \psi_\downarrow (\tau) \; ,
\label{Gl5}
\end{eqnarray}
where $Z_\alpha$ is the partition function, and
 $N^0(\epsilon)$ is the density of states (DOS) of the non--interacting 
electrons.
As the results do not much depend on its precise form
we choose a half-elliptic DOS $N^0(\epsilon) = 
[(2t^*)^2 - \epsilon^2 ]^{1/2}/(2 \pi t^{*2})$.
The constraint ${\mathbf m}_{st} \parallel {\mathbf H}$ is enforced
by set\-ting  the off-diagonal (in spin space) elements
of the Green function equal to zero: $G^{\downarrow \uparrow}_\alpha = 
-< \psi_\alpha^\downarrow \psi_\alpha^{\uparrow *} > \equiv 0$.
From now on $t^* \equiv 1$ will set our energy scale,
i.e.~the total band width of $N^0(\epsilon)$ is equal to 4.
The Dyson equation (\ref{dys})  introduces the lattice into the problem.
It couples $A-$ and $B-$sublattices and can be solved by a simple
integration, even analytically for the above DOS. 
The functional
integral (\ref{fi}) however is highly non--trivial since it couples all 
Matsubara frequencies.
Georges and Kotliar \cite{Georges92} and Jarrell \cite{Jarrell92} 
realized that the action (\ref{Gl5}) 
is equivalent to that of an Anderson impurity model, and can therefore
be treated by standard techniques developed for this model.
Here we employ a finite temperature, auxiliary field
Quantum Monte Carlo (QMC) method \cite{Hir86,Ulm95}. 
In this approach the electron--electron interaction 
is formally replaced by an interaction of independent electrons 
with a dynamical, auxiliary field of Ising--type spins.
To this end the interval $[0,\beta]$
is discretized into $\Lambda$ steps of size $\Delta\tau=\beta/\Lambda$.  
Equivalently, there is a high energy cut--off of Matsubara frequencies, 
i.e.~$|\omega_n|=\pi T |2n+1| <\pi/\Delta\tau$, 
$n=-\Lambda/2,\dots, \Lambda/2-1$.
All quantities have to be extrapolated to $\Delta\tau\to 0$.
The computer time grows like $\Lambda^3\propto\beta^3$, restricting 
$\Lambda$ to values below $\sim 150$ and $\beta\le 50\dots 70$
on present supercomputers. 
For small $\Lambda$ $(\Lambda\leq 20)$ one can perform a full enumeration
(instead of the Monte Carlo sampling) of all $2^\Lambda$ possible 
configurations of the auxiliary field.
We never encountered a minus--sign problem, hence no further approximations
(like the ``fix--node'' method) were necessary. 
 
The self--consistency is obtained iteratively as follows: the Green function 
$G$ (omitting indices) is calculated from some initial self energy, 
e.g.~$\Sigma\equiv 0$, by the Dyson equation (\ref{dys}).
Now the new Green function $G_{\rm new}$ is determined by
solving (\ref{fi}) with the QMC method. Finally, the calculation of the new
self energy $\Sigma_{\rm new}=\Sigma-G^{-1}_{\rm new}+G^{-1}$
completes one iteration. To improve convergence in the symmetry broken case
$G_A$ and $G_B$ are updated by the Dyson equation  (\ref{dys})
after every QMC simulation for one sublattice (Eq.~\ref{Gl4}).
In the symmetry broken phases, typically $10-20$ iterations with 20000 MC 
sweeps are 
necessary to obtain a convergence of $\sim 10^{-3}$. The calculation of 
a magnetization curve at $\beta=50$
takes about 100 hours on a Cray-Y-MP.
Close to a phase transition the convergence is much slower
and the statistical errors are  larger due to strong fluctuations,
in particular in the case of a second order phase transition. 
These effects limit the accuracy in the determination of the critical values
of the model parameters, e.g.~the critical magnetic field (see Sec.~\ref{ic}).
At large $U$-values $(U>4)$ the Monte Carlo sampling becomes more and
more inefficient due to ``sticking'' problems, i.e.~there are two (or more)
minima in the free energy and the single spin--flip algorithm is no longer able
to transfer between them.

From the resulting Green functions we calculate the densities and
the homogeneous and 
antiferromagnetic magnetization:
\begin{eqnarray}
n_{\alpha \sigma} &=& 1 + T \sum_n G^\sigma_{\alpha n},\nonumber\\
m & = & \frac{1}{2} \sum_{\alpha \sigma} \sigma n_{\alpha \sigma},
\nonumber \\
m_{st} & = & \frac{1}{2} \sum_{\alpha \sigma} \alpha \sigma n_{\alpha
  \sigma}.
\end{eqnarray}

Since we are  interested in the question whether 
the system is insulating or metallic we also
determine the electronic compressibility 
$\kappa_e \equiv \partial n/\partial \mu$. It was calculated both
by numerical differentiation of $n(\mu)$
and from the two--particle correlation functions
(see \cite{Ulm95} and Appendix \ref{sus}). 
Both results agree within the statistical errors:
however, the latter method is much more time consuming.

\section{Results for half filling}
\label{resn1}
\label{reshf}

\subsection{Weak coupling}
\label{wc}
For weak coupling ($U \ll t$) we expect the Hartree--Fock
approximation
to give an, at least qualitatively\cite{pvdwc},
correct picture of the metamagnetic phase transitions, especially of
their order.
Within this {\it static} mean--field theory  electronic correlations
are neglected and the interaction is decoupled
 as
\begin{eqnarray}
   \hat{n}_{i \sigma}  \hat{n}_{i - {\sigma}} &\stackrel{\mathrm {HF}}
   {\longrightarrow}&
  \hat{n}_{i \sigma} <\! \hat{n}_{i -{\sigma}}\! > +  <\! \hat{n}_{i \sigma}\! >
  \hat{n}_{i   -{\sigma}} \nonumber \\
  & &- <\! \hat{n}_{i -{\sigma}}\! >  <\! \hat{n}_{i \sigma}\! > .
\label{dec}
\end{eqnarray}
Note, that while for ${\mathbf e} \parallel {\mathbf m}_{st} \parallel 
{\mathbf m}$ the Fock term vanishes,
$ <\! \hat{c}_{i \uparrow}^\dagger \hat{c}_{i \downarrow}^{\phantom{\dagger}} 
\! > = 0$,
its presence is essential in
the  case ${\mathbf m}_{st} \perp {\mathbf m}$
(see Appendix \ref{perpo}). We confine our investigations to states
with homogeneous sublattice magnetization
as described by the Ansatz
\begin{equation}
  <\!n_{i \in \alpha, \sigma}\!> = \frac {1}{2} ( n  + \sigma \; m +
  \alpha  \; \sigma \; m_{st}) .
\label{hde}
\end{equation}
Applying this Hartree--Fock decoupling scheme one obtains 
the effective, one--particle Hamiltonian 
\begin{eqnarray}
 {\cal H}_{\mathrm {HF}} & =  & 
  \sum_{N N \sigma} t_{i j} \hat{c}_{i \sigma}^\dagger \hat{c}_{j
   \sigma} + \nonumber \\ && 
  \sum_{\alpha \; i \in \alpha \; \sigma} \frac {U}{2} (n- \sigma
  m - \sigma \alpha m_{st})  \hat{n}_{i \sigma} - ( \mu + \sigma H )
  \hat{n}_{i \sigma}  \nonumber \\
  &&
  - \frac{1}{2} \frac{U}{4} \sum_{\alpha \; i \in \alpha \; \sigma} 
\left[ n^2- (m+\alpha  m_{st})^2 \right] .
 \label{hhe}
\end{eqnarray}
This Hamiltonian is diagonalized and the 
one--particle energies $\tilde{\epsilon}_\sigma$
are calculated as
\begin{equation}
\tilde{\epsilon}_\sigma = 
  {\mathrm {sgn}}(\epsilon)  \sqrt{\epsilon^2+\left(\frac{U}{2} m_{st}\right)^2}
   - \sigma \left(\frac{U}{2}m +H\right),
\label{hfe}
\end{equation}
where  ${\mathrm {sgn}}(\epsilon)$ denotes the sign  of the
non--interacting electron energy $\epsilon$.
In the antiferromagnetic phase the DOS has a gap of
 width $U m_{st}$  with square root singularities at its edge.

From the one--particle energies (\ref{hfe}) 
the grand potential per lattice site ($L$ being the number of lattice sites) 
is obtained directly as
($\tilde{\mu}= \mu - \frac{1}{2} U n$)
\begin{eqnarray}
\Omega/L  & = & -\frac{ 1} {\beta}  \ln {\cal Z} \nonumber\\
         &  = & -\frac{ 1} {\beta} \sum_{\sigma} \int  {\mathrm d} \epsilon
            N^0(\epsilon)
             \ln \! \left( 1 + e^{-\beta ( \tilde{\epsilon}_\sigma -
               \tilde{\mu})} \right) 
      \nonumber    \\&& +  \frac{U}{4} (m_{st}^2 +m^2)-\frac{U n^2}{4} . 
\label{ome}
\end{eqnarray}
The potential $\Omega$
has two shallow minima, one corresponding to the 
paramagnetic state ($m \neq 0, m_{st} = 0$) and the other to 
the antiferromagnetic
state  ($m \approx 0, m_{st} \neq 0$). 
By applying a sufficiently strong magnetic field to the
antiferromagnetic state, the 
paramagnetic minimum becomes the lowest, such that 
a first order phase transition takes place.
The reason for the occurrence of a 
first order phase transition is that, within the
(static) Hartree--Fock approximation, the $U-$term  becomes
minimal for largest (static) local moments. Therefore
pure antiferromagnetic and ferromagnetic order are both energetically
favored. Mixed states with $m \neq 0$  and $m_{st} \neq 0$ 
-- which would occur in the case of second order phase transitions --
have a small local moment, i.e.~a high Hartree--Fock energy, 
on every second site.

To calculate the magnetization curves the self--con\-sistent
Hartree--Fock equations are obtained from the minimization conditions
${\partial \Omega}/{\partial m_{st}}  = 0$ and
${\partial \Omega}/{\partial m} = 0$:
\begin{eqnarray}
  m_{st} &=& \frac{U}{2} \sum_\sigma \int {\mathrm d} \epsilon N^0(\epsilon)
  \frac{- m_{st}} {\tilde{\epsilon}_\sigma} 
{\frac{1}{1 + e^{\beta \left(\tilde{\epsilon}_\sigma-\tilde{\mu}\right)}}}
\label{sel1}\\
  m &=& \sum_\sigma \int {\mathrm d}  \epsilon N^0(\epsilon)
 \;\sigma \;   {\frac{1}{1 + e^{\beta \left(\tilde{\epsilon}_\sigma
       -\tilde{\mu}\right)}}}.
\label{sel2}
\end{eqnarray}
These equations are solved  numerically by iteration and integration
according to  Newton--Cotes rules.

\begin{figure}[t]
\unitlength1cm \epsfxsize=8.6cm   \epsfbox{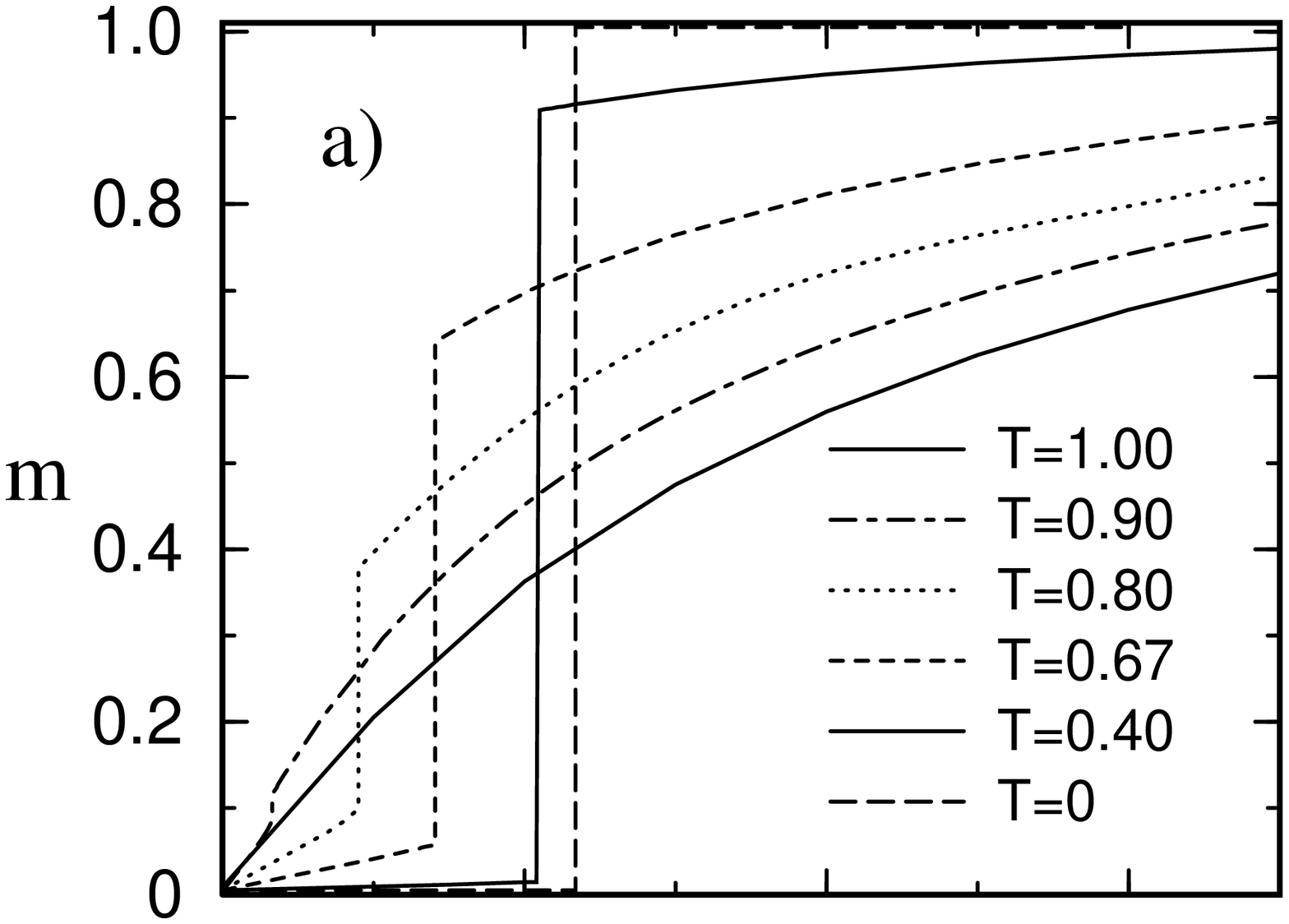}
\vspace{-1.81cm}

\unitlength1cm \epsfxsize=8.6cm   \epsfbox{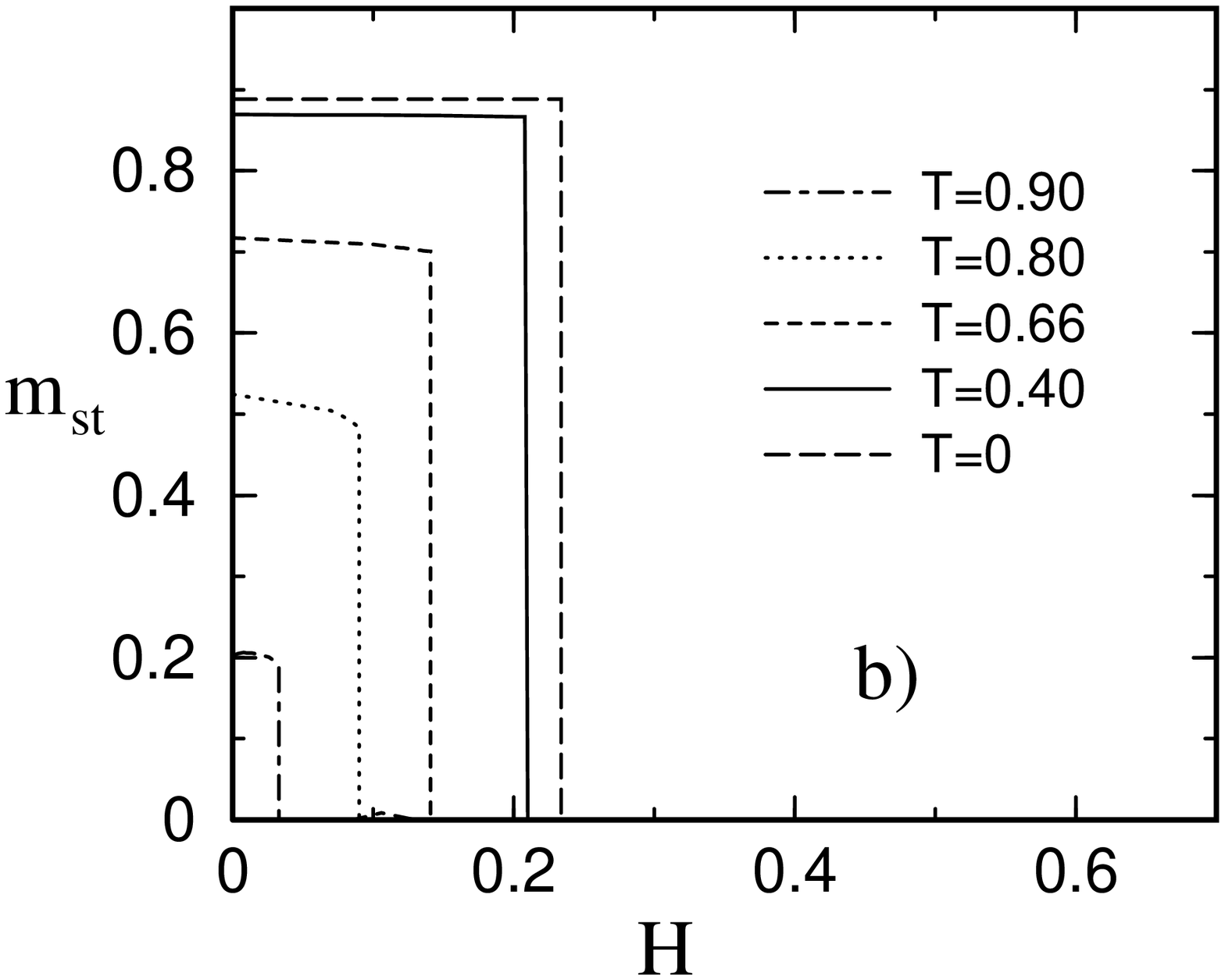}
\caption{ a) Magnetization $m$ vs.~magnetic field $H$
 in Hartree--Fock approximation for $U=4$
at different temperatures $T$ showing metamagnetic behavior.
b) Order  parameter for the  
metamagnetic phase transition (the staggered  magnetization) $m_{st}$ vs.~$H$.
The first order phase transition is clearly seen.
\label{hfm}}
\end{figure}

\begin{figure}
\unitlength1cm \epsfxsize=8.6cm   \epsfbox{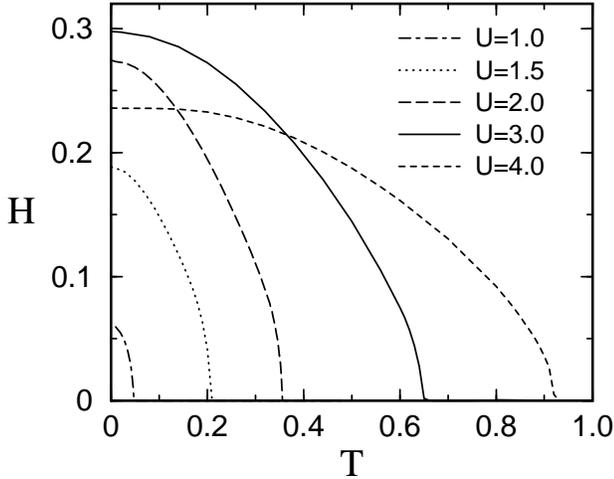}
\caption{$H-T$ phase diagram for different values of $U$ in 
Hartree--Fock approximation.
All phase transitions are of first order. 
Below the curves the antiferromagnetic phase is stable.
\label{hfp}\\}
\end{figure}

Within the Hartree--Fock approximation the meta\-magnetic
phase transition is found to be of {\em first} order for 
all $T$, even for $U=4$ ( = band width);
see Fig.~\ref{hfm} and \ref{hfp}.  Hence a tricritical point never occurs. 
In this parameter range the  
Quantum Monte Carlo calculations, however,
already show second order transitions
in a broad range of temperatures (see Sec.~\ref{ic}).
Hence the Hartree--Fock solution can neither describe
the experimental situation, where tricritical points are known to occur, nor 
the correct behavior of the model
for intermediate values of $U$.

To estimate the anisotropy energy
associated with  the easy axis we compare the Hartree--Fock
energies of the configurations with
${\mathbf m}_{st} \parallel {\mathbf m}$ and ${\mathbf m}_{st} 
\perp {\mathbf m}$
(for details see Appendix \ref{perpo}).
At half filling and for $U$ equal to the band width, the difference
between the free energy of these configurations 
 does not exceed a few 
percent of the band width, i.e.~${\cal O} (10^{-2} eV)$.
In this situation the spin--orbit interaction, which can be relatively
strong, ${\cal O} (10^{-1} eV)$, indeed leads to
a strong anisotropy, i.e.
an easy axis $\mathbf{e}$, along which ${\mathbf m}_{st}$ is rigidly
fixed.

Metamagnetic phase transitions in itinerant, metallic systems
where hitherto described by the theory of
``itinerant electron metamagnetism'' (IEM).
In the case of an antiferromagnetic system in a magnetic field
Moriya and Usami\cite{Moriya77} proposed a Landau
theory with free energy
\begin{eqnarray}
F(m,m_{st}) &&=\frac{1}{2 \chi_m} m^2 + \frac{1}{2 \chi_{st}} m_{st}^2 +
a \;  m^4 +a' \; m_{st}^4
 \nonumber \\ 
&&+b \;  m^2 m_{st}^2 +
b' \;  ({\mathbf m}\!\cdot\! 
{\mathbf m}_{st})^2 - H m,
\label{iemlandau}
\end{eqnarray}
where $\chi_m$ and $\chi_{st}$ are the homogeneous and staggered 
susceptibility, respectively, and the coefficients $a$, $a'$, $b$, 
and $b'$ are the fourth--order derivatives of the non-interacting free energy.
Within the IEM theory 
the Coulomb interaction is treated in random phase
approximation. The corresponding  
 susceptibilities are given as 
\begin{eqnarray}
 \frac{1}{ \chi^{\phantom{0}}_{m}}=  \frac{1}{ \chi^0_{m}} - U, 
& \hspace{2em} &
\frac{1}{ \chi^{\phantom{0}}_{st}}=  \frac{1}{ \chi^0_{st}} - U,
\label{RPA}
\end{eqnarray}
where $\chi^0_{m}$ and $\chi^0_{st}$ are the respective susceptibilities 
of the non-interacting system.
The random phase approximation 
for these susceptibilities is equivalent to the Hartree--Fock scheme 
described above.
Therefore we may ask whether we obtain the IEM in the limit
$U\ll t$ (where $m, m_{st} \ll 1$). The answer is not straightforward since 
the prefactors in the expansion (\ref{iemlandau}) depend, for example, 
on the lattice structure. On bipartite lattices, as discussed here, 
they diverge for $T \rightarrow 0$.
Thus an expansion of the free energy in powers of $m_{st}$ and $m$
as assumed in the  IEM Landau theory is not possible in general
(for details see Appendix \ref{icmhf}).

\subsection{Strong Coupling}
\label{sc}
In the 
limit $U \gg t$ the Hubbard model at half filling ($n = 1$)
is equivalent to an effective Heisenberg spin--model
\begin{equation}
\hat{H}_{\mathrm Heis} = \frac{J}{2} \sum_{NN}  {\mathbf{\hat{S}}}_i \cdot
{\mathbf{\hat{S}}}_j - 2 H \sum_i \hat{S}_i^z,
\label{isiaf}
\end{equation}
where the antiferromagnetic exchange coupling
is obtained in second order perturbation theory as
$J = 4t^2/U$. Spin operators are defined as 
$\hat{S}_i^z = \frac{1}{2}( \hat{n}_{i \uparrow}- \hat{n}_{i \downarrow})$, 
$\hat{S}^x_i = \frac{1}{2}(
\hat{c}_{i \uparrow}^\dagger \hat{c}_{i \downarrow}^{\phantom{\dagger}}
+\hat{c}_{i \downarrow}^\dagger \hat{c}_{i \uparrow}^{\phantom{\dagger}})$, and 
$\hat{S}^y_i = -\frac{i}{2}(
\hat{c}_{i \uparrow}^\dagger \hat{c}_{i \downarrow}^{\phantom{\dagger}} 
- \hat{c}_{i \downarrow}^\dagger \hat{c}_{i \uparrow}^{\phantom{\dagger}})$. 
For this model the 
Weiss molecular field theory becomes exact
in $d = \infty$ yielding, 
under the constraint of uniaxial magnetization,
the same results as for the Ising model.
For Ising models
 metamagnetic phase transitions 
are well-studied \cite{7}. In the case of a purely
antiferromagnetic next-neighbor coupling  (see (\ref{isiaf}))
the phase transitions are
of first order only at $T=0$, but of {\em second} order 
at all $T>0$.
The transition line in the $H-T$ phase diagram has indeed the form shown in
Fig.~1a, { but with}
a tricritical temperature of $T_t = 0$. 
This behavior can be understood already within Weiss molecular 
field theory, where
the ground state energy per site is
\begin{equation}
E(m,m_{st}) = \frac{J^*}{8} (m^2-m_{st}^2) - H m,
\end{equation}
with
\begin{equation}
J^* = Z J = \frac{4{t^*}^2}{U}.
\end{equation}
Minimization with respect to $m$ and $m_{st}$ shows that
the fully polarized antiferromagnet ($m_{st}=1$) has 
lowest energy 
for $H < J^*/4$, whereas the fully polarized ferromagnet
($m=1$) is energetically favored for $H > J^*/4$.
Thus, by applying a magnetic field a first order transition is induced.
At   $H = J^*/4$ the states are highly degenerated since {all} magnetic phases
with $m+m_{st}=1$ have the {same} energy. For $T>0$
this degeneracy is lifted by entropy which disfavors
fully polarized phases. Therefore the first order
transition at $T=0$ immediately becomes second order for $T>0$, 
i.e.~$T_t=0$.
Indeed, a tricritical point at a {\em finite} temperature is only obtained
in the case of spin interactions which simultaneously favor both
fully polarized antiferromagnetic and ferromagnetic configuration.
In particular, adding 
a ferromagnetic interaction $J'$  between next nearest neighbors (NNN)
on a simple cubic lattice as in (\ref{isi})
stabilizes both ferro- and antiferromagnetic order \cite{7}.

While in the case of effective spin models
a ferromagnetic NNN coupling term is introduced {\em ad hoc},
simply to obtain the first order phase 
transition, this term naturally arises if we expand the strong
coupling perturbation series of the Hubbard model
to ${\cal O} (t^4/U^3)$. However, besides this $J'$-Term there also appear
additional {\em four}--spin terms.
For the hypercubic lattice the effective Hamiltonian
$H_{\mathrm eff}$ reads\cite{pvdsc}
\begin{eqnarray}
H_{\mathrm eff} &=& - \frac{J}{4} \sum_{N N}Q_{i j} +  
                   \frac{J'}{4}\sum_{i, \tau' \neq \pm \tau} 
                       Q_{i + \tau', i+\tau} \nonumber \\
                  && +\frac{t^4}{U^3} A
                   \sum_{ \{\Box\}} \left( Q_{12} Q_{34} +  Q_{14} Q_{23} -
                      Q_{13} Q_{24} \right) ,
                    \label{Hsc} \\
J &=& 4 \left( \frac{t^2}{U}+ B\frac{Z t^4}{U^3}\right)\\
J' &=& 4 \frac{t^4}{U^3} C.
\end{eqnarray}
Here $\tau$ and $\tau'$ are lattice vectors connecting a site to its
$Z$ neighbors, and $\Box$ represents a plaquette. Each plaquette is
counted only once: the four sites $\{1,2,3,4\}$ represent its four
corners in clockwise or anticlockwise order.
The constants $A$, $B$ and $C$ depend on the lattice, and
the Hermitian operators $Q_{ij}$ are defined as
\begin{equation}
 Q_{i j} = -2 \left( {\mathbf {\hat{S}}}_i \cdot {\mathbf {\hat{S}}}_j - 
\frac{1}{4}\right).
\end{equation}

The plaquette contribution competes with the ferromagnetic NNN term ($J'$) and
  drives the system to second order phase
 transitions.
For the hypercubic lattice the plaquette contribution
is stronger than the ferromagnetic NNN term, yielding second order
phase transitions even for $T=0$. The same is true for
the Bethe lattice where, in fact, $J'<0$ (for details see Appendix \ref{asc}).
Thus in strong coupling perturbation theory the metamagnetic phase 
transition is of {\em second order even at} $T=0$.

\subsection{Intermediate coupling} 
\label{ic}
\begin{figure}[t]
\unitlength1cm \epsfxsize=8.6cm   \epsfbox{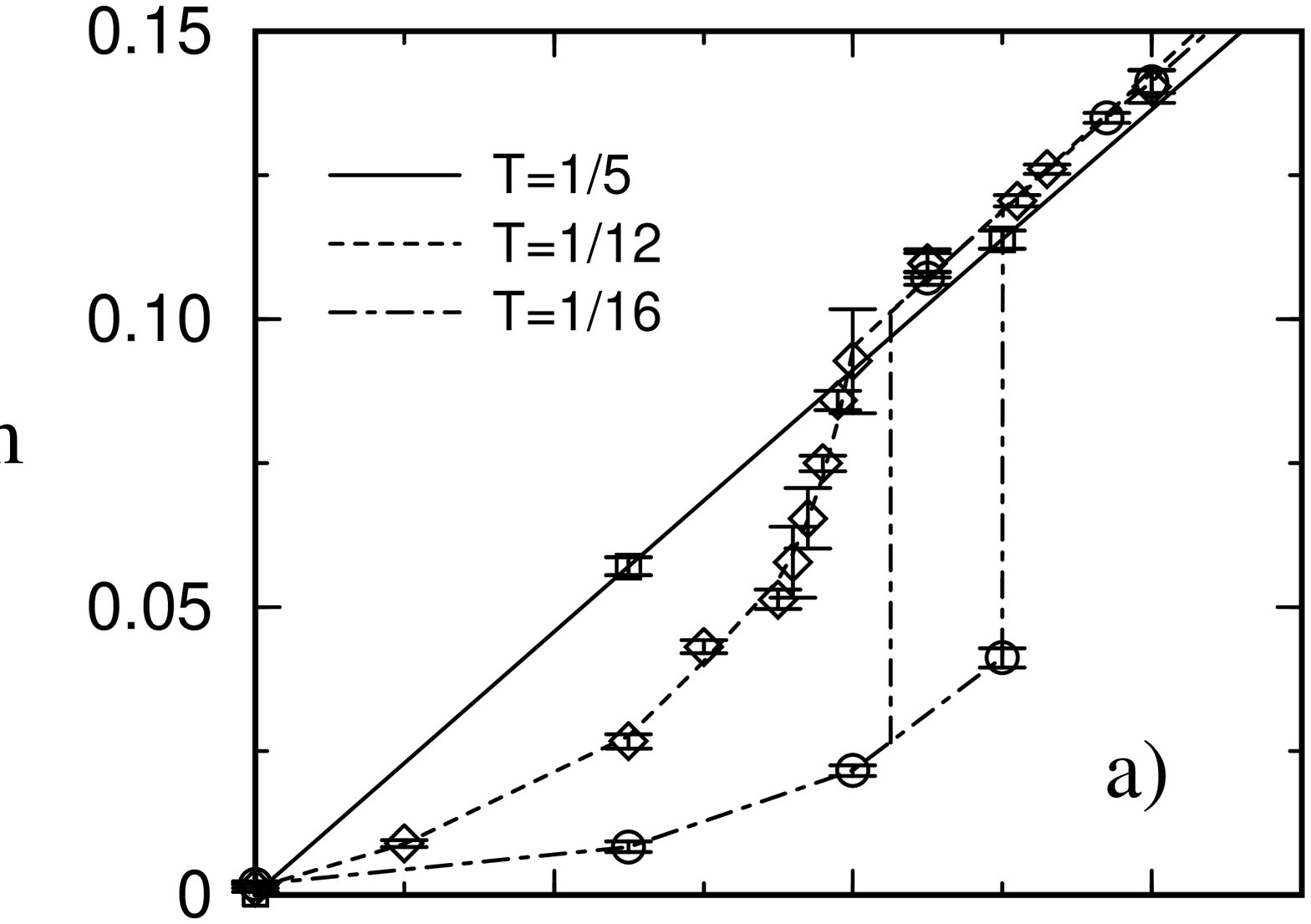}
\vspace{-1.81cm}

\unitlength1cm \epsfxsize=8.6cm   \epsfbox{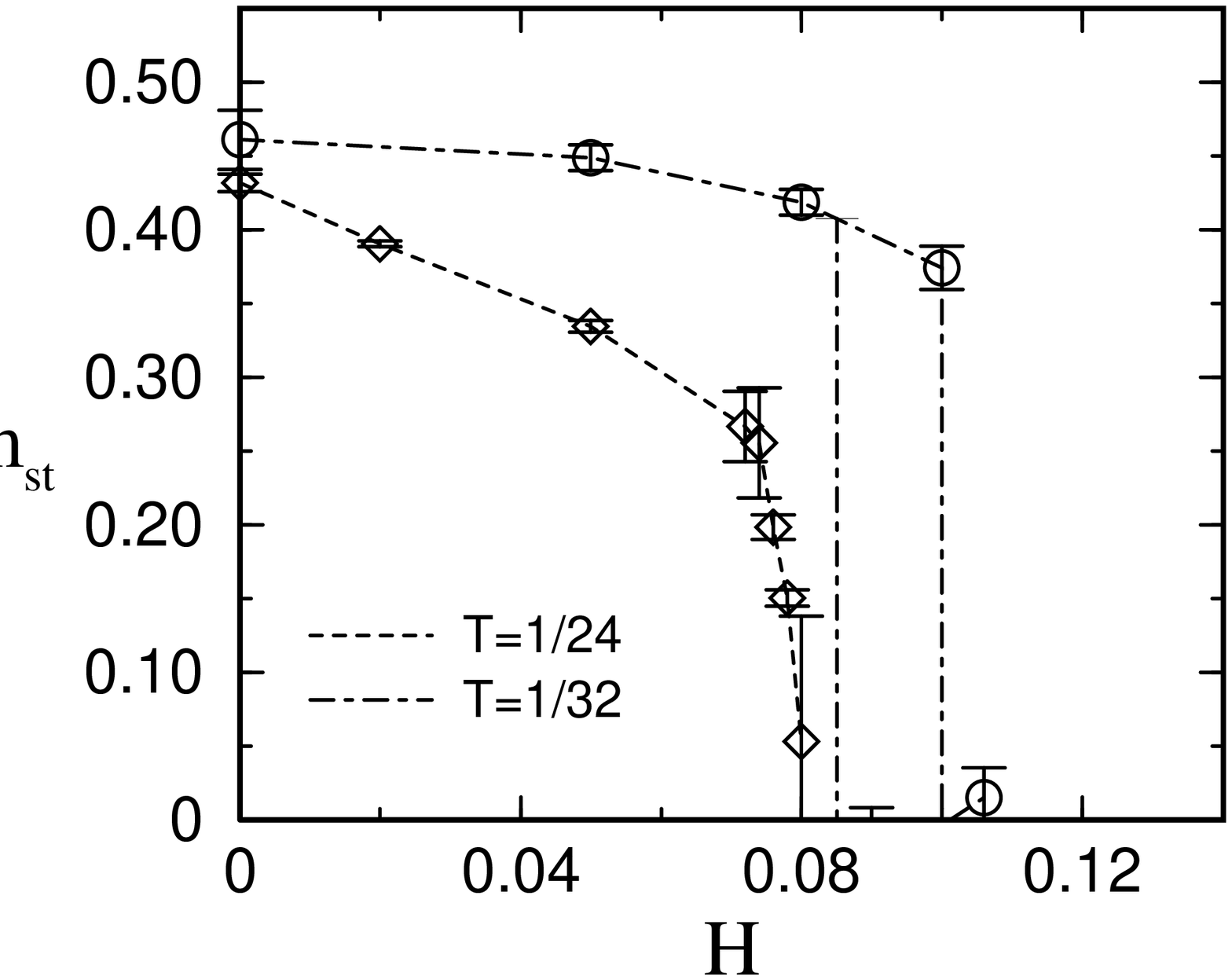}
\caption{QMC results, including error bars, a) for the magnetization
 $m(H)$
as obtained for the $d=\infty$ Hubbard model with easy axis at half filling
and $U=2$ (= half band width) for different temperatures. 
b) Staggered magnetization $m_{st}(H)$ for the two temperatures below $T_N$.
\label{mvsH}\\}
\end{figure}

The perturbation analysis described above demonstrates that the order of the
metamagnetic phase transition depends on the Coulomb interaction $U$
in a delicate way.
For small $U$ the phase transition is purely of first order
and for large $U$ of second order.
Apparently, the tri- or multicritical point linking these two regimes
must be found at intermediate coupling.
In this important, non--perturbative regime 
Quantum Monte Carlo techniques are employed to solve the problem numerically
without any further approximation.
The results for the magnetization $m(H)$ and the staggered magnetization 
$m_{st}(H)$ are shown in Fig.~\ref{mvsH}   for $U=2$. 
Below the N\'eel temperature a 
metamagnetic behavior is clearly seen: for small magnetic fields the 
magnetization is exponentially suppressed with temperature. Then, towards 
the metamagnetic phase transition, the susceptibility increases drastically 
and becomes maximal at the critical field $H_c$. Second order phase transitions
are observed for $1/14\leq T\leq T_N=0.114 \pm 0.006$, 
whereas the transition is of first order at
lower temperatures, i.e.~$T\leq 1/16$.
At the phase transition the order parameter, i.e.~the staggered magnetization,
vanishes. From the curve $m_{st}(H)$ the critical field $H_c$ and also the
order of the phase transition is determined by a square root fit for second 
order transitions
and by the
mean of the hysteresis for first order transitions.
 
\begin{figure}[t]
\unitlength1cm \epsfxsize=8.6cm   \epsfbox{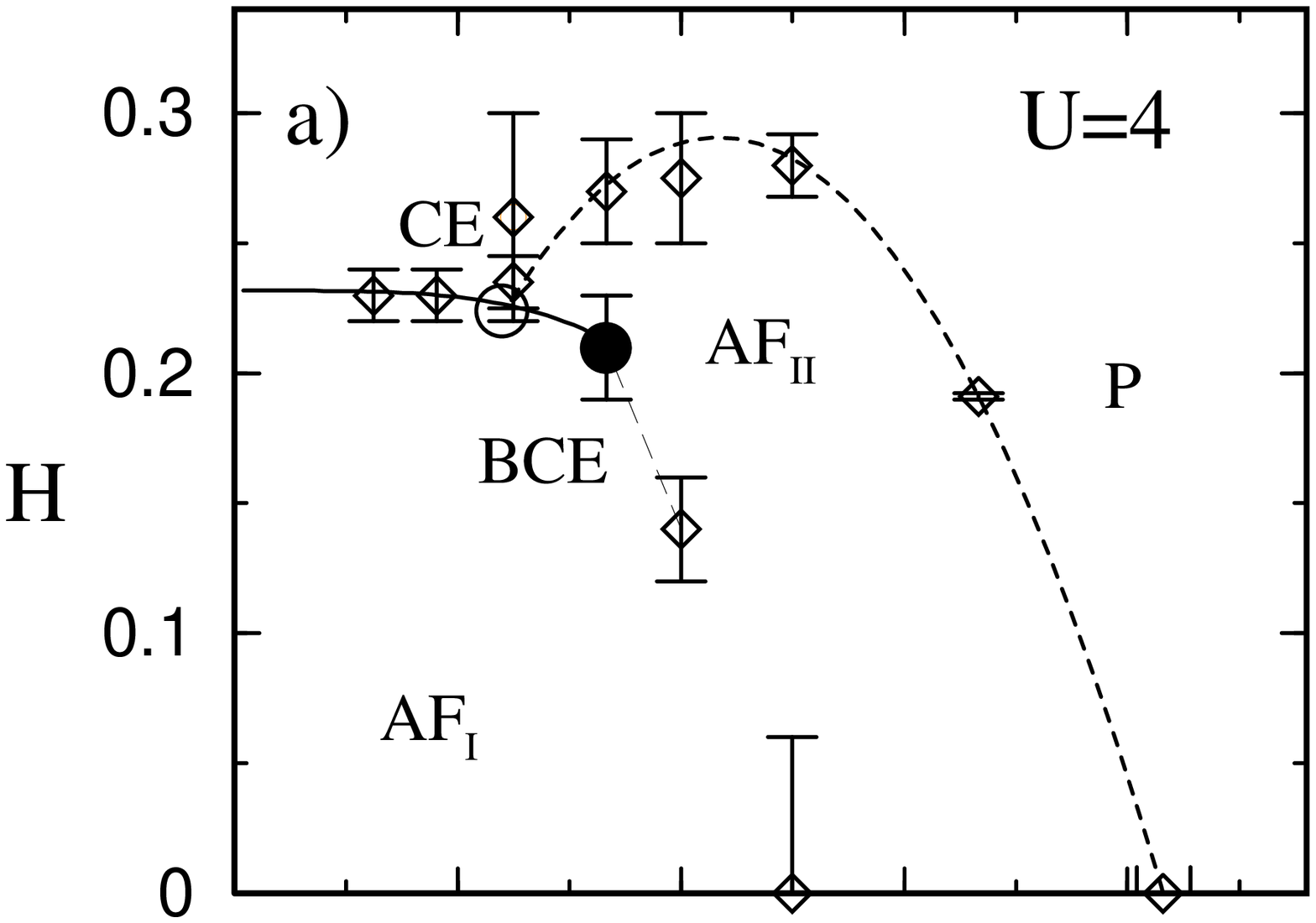}
\vspace{-1.81cm}

\unitlength1cm \epsfxsize=8.6cm   \epsfbox{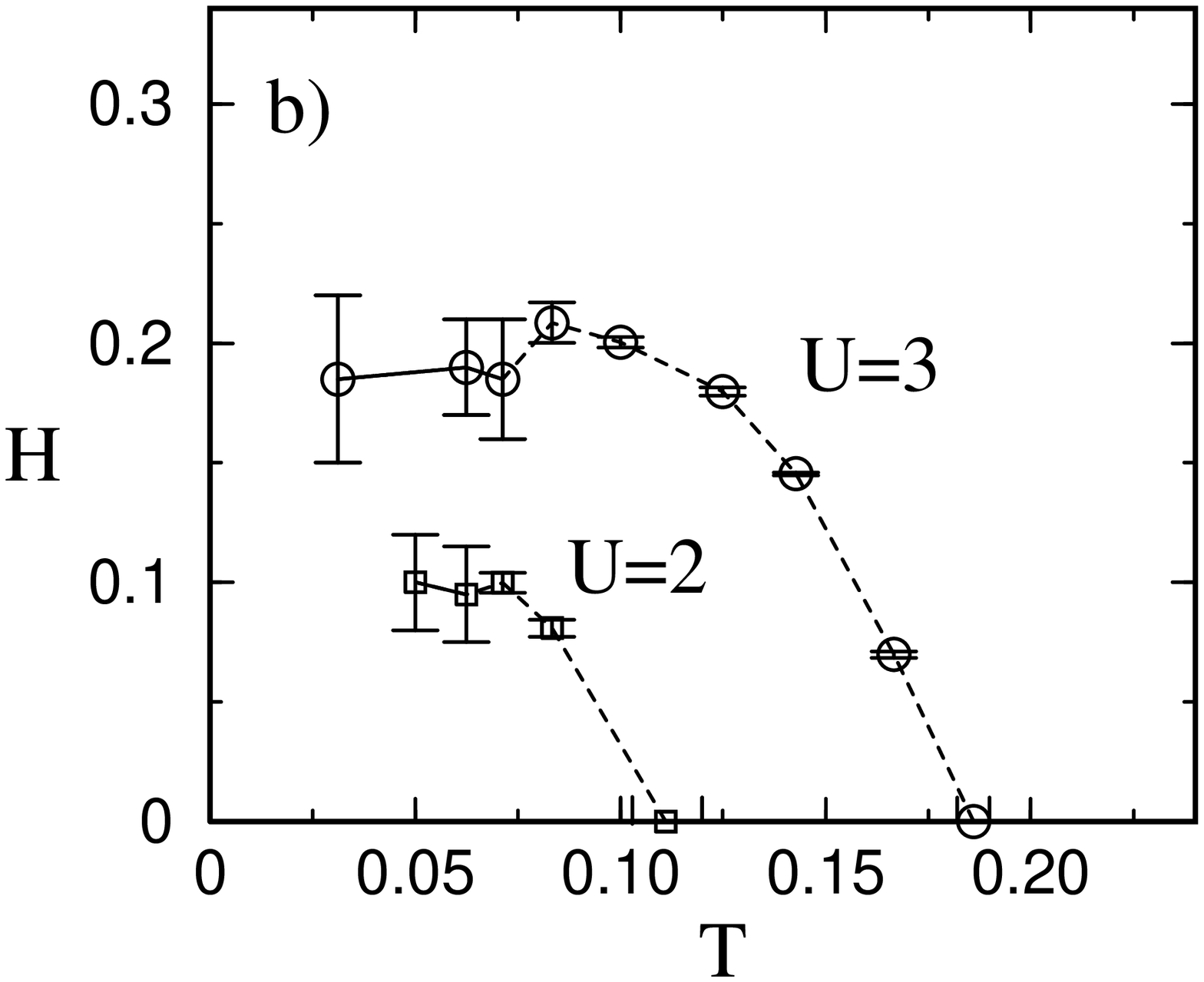}
\caption{$H-T$ phase diagram for the $d=\infty$ Hubbard model with 
easy 
axis at half filling
as constructed from the 
QMC results for $m(H)$ and $m_{st}(H)$,
a) $U=4$, b) $U=2,3$. Second order phase
transitions are indicated by dashed lines, first
order transitions by solid lines. Curves are guides to the eye only.
\label{pd}\\}
\end{figure}

Using these values of $H_c$ the phase diagrams,  Fig.~\ref{pd}, 
for different values of $U$ are constructed. 
The case $U=4$ ( = band width) and half filling, Fig.~\ref{pd}a,
was already presented in Ref.~\cite{Hel96}. 
This phase diagram  shows {both}
 first order (for $T < 1/16$) {\em and}
 second order phase transitions
(for $1/8 < T < T_{\mathrm {N}} \approx 0.2$).
The field dependence at intermediate temperatures, i.e.~$1/16 < T <
1/8$, is more complex:
 $m_{st}$ is almost field-independent in region AF$\mathrm _I$ 
(where $m \simeq 0)$, decreases sharply at the boundary to region 
AF$\mathrm _{II}$ (where  $m > 0$) and, upon further increase of the field, 
vanishes in a second order transition.
Although the error bars do not permit an unambiguous interpretation
it hence seems that the order parameter decreases by {\it two} consecutive
transitions: the first one being of first order or 
corresponding to an
anomaly, and the second one being of second order. 
Taken together  the results seem to correspond to the
scenario of Fig.~\ref{scheme}b.
For $U=2$ 
phase transitions of first order are found for $T \leq 1/16$ 
and of second order
for $ 1/14 \leq T < T_{\mathrm N}=0.114 \pm 0.006$ (see Fig.~\ref{pd}b).
Here, the temperature regime with two consecutive 
transitions, as obtained for $U=4$, has disappeared or has become very small:
the scenario is similar to Fig.~\ref{scheme}a.
The phase diagram for $U=3$, also displayed in Fig.~\ref{pd}b,
shows some features of the phase diagram
in Fig.~\ref{scheme}b or $U=4$, respectively. In particular one observes a
maximum in the second order phase transition line and
different slopes for the second and first order line at the crossover point.
However, the
numerical data do not indicate the existence of two consecutive transitions.
Thus this phase diagram lies in between the scenarios depicted in 
 Figs.~\ref{scheme}a and \ref{scheme}b.

\begin{figure}
\unitlength1cm \epsfxsize=8.6cm   \epsfbox{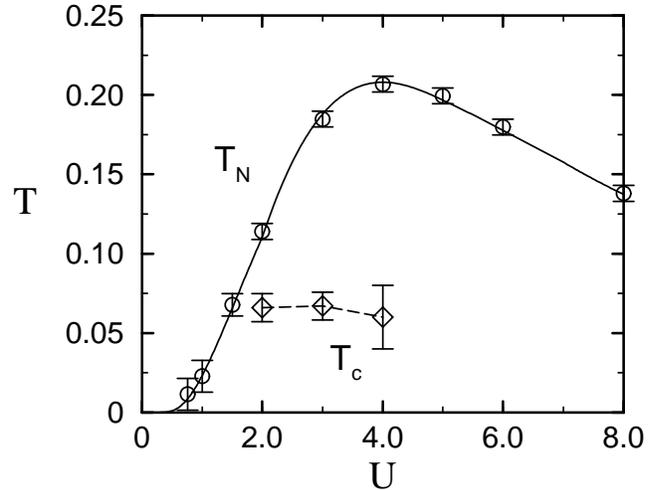}
\caption{N\'eel temperature $T_N$ (circles) and (tri-)critical 
temperature $T_{c}$ (diamonds)
vs.~$U$. Above  $T_N$ the system is paramagnetic.
In an external magnetic field
 the order parameter vanishes 
in a second order metamagnetic phase transition 
 for $T_c<T<T_{N}$ and in a first order
transition for $T<T_{c}$, respectively. 
\label{TcvsU}\\}
\end{figure}

To study the influence of the interaction $U$ qualitatively 
and quantitatively,
it would be desirable to calculate $H-T$ phase diagrams 
at even larger values of $U$.
Unfortunately the Quantum Monte Carlo
approach fails in this case due to the problem of ``sticking''
as mentioned in Sec.~\ref{dmft}.
Therefore, we concentrate on the 
crossover from the intermediate coupling regime
with first {\em and} second order transitions to the weak coupling
regime with first order transitions {\em only}. 
The results for half filling are collected in Fig.~\ref{TcvsU} 
showing the $U$--dependence of two transition temperatures:
the top curve is the N\'eel temperature $T_{\mathrm {N}}$ at $H = 0$  
(taken from Ref.~\cite{Ulm95}).
The lower curve corresponds to  the temperature $T_{c}$ 
where the second order phase transition
line terminates, i.e.~it represents either the tricritical or the 
critical temperature  of Fig.\ref{scheme}.
For temperatures below $T_{c}$ a first order
meta\-magnetic phase transition is observed in an external magnetic field.
Fig.~\ref{TcvsU} reveals 
the crossover from intermediate coupling with
first and second order phase transitions 
to weak coupling with first order transitions only:
as $U$ decreases the regime with second order 
phase transitions ($T_{c} < T < T_N$) shrinks, while the temperature 
regime for first order transitions remains nearly unchanged up to 
$U \approx 2$.

\section{Results away from half filling}
\label{resbhf}
\begin{figure}
\unitlength1cm \epsfxsize=8.6cm   \epsfbox{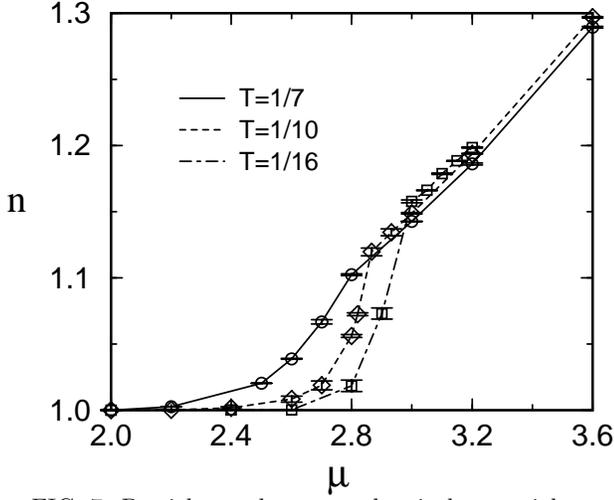}
\caption{Particle number $n$ vs.~chemical potential $\mu$ as 
calculated by grand 
canonical QMC simulations in the antiferromagnetic phase for $U=4$.
\label{phassep}\\}
\end{figure}

In the preceding sections metamagnetic transitions were investigated in 
the case of half filling.
Beyond half filling 
the commensurate antiferromagnetic phase 
remains stable in the parameter regime under consideration (
$\delta = |1-n| \leq 0.075$, $T\geq 1/32$).
Incommensurate spin density waves become stable only 
in a small density regime at lower temperatures
\cite{Freericks95}. Another possible instability of the antiferromagnetic 
phase away from half filling is phase separation, found 
within the Hartree--Fock approximation and in second order perturbation 
theory at constant order parameter at T=0 \cite{pvd95}. However,
at least for $T \geq 1/16$
we do not observe phase separation 
since the electronic compressibility $\kappa_e = \partial n/\partial \mu$ 
is finite and positive 
(see Fig.~\ref{phassep}).

\begin{figure}[t]
\unitlength1cm \epsfxsize=8.6cm   \epsfbox{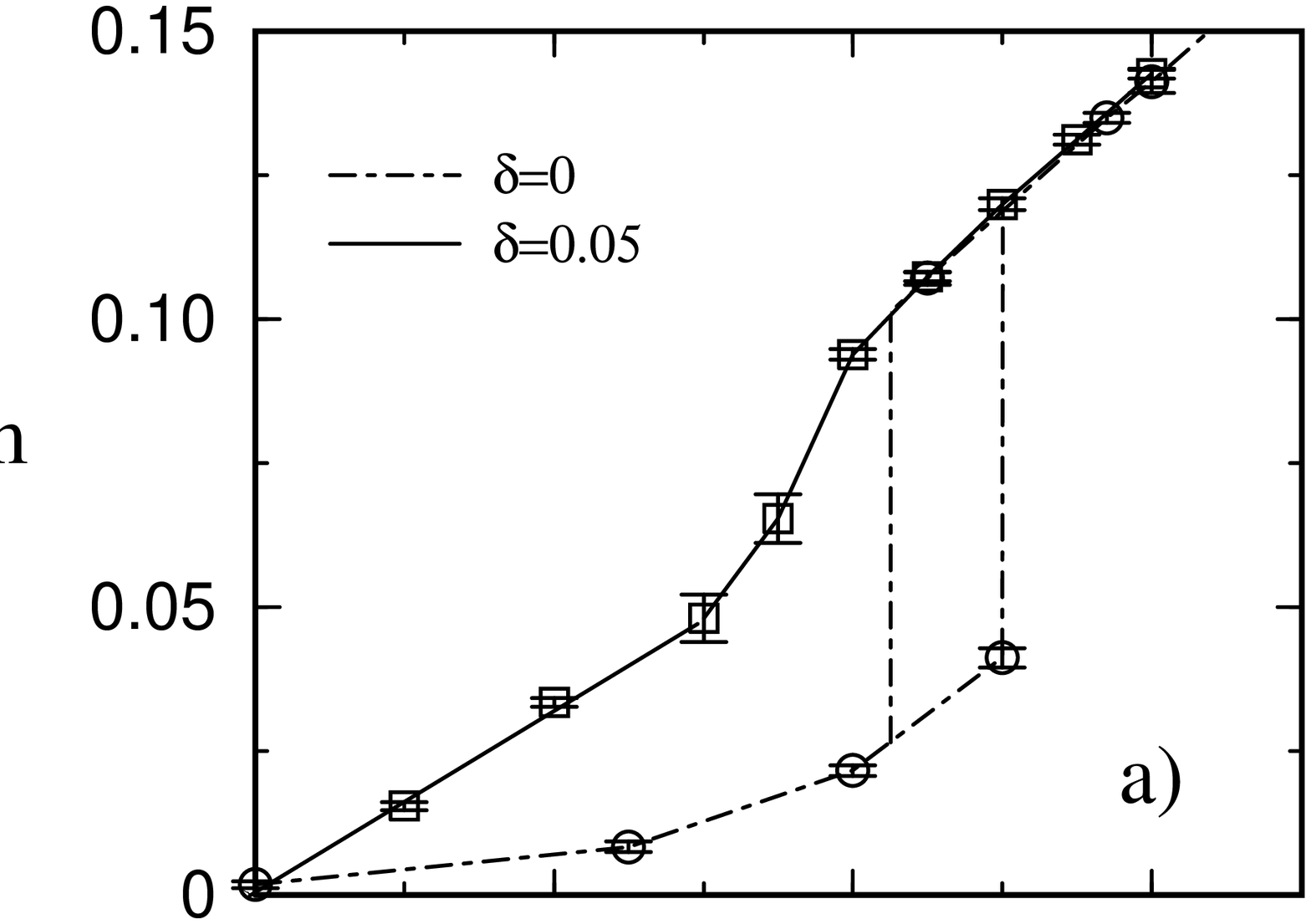}
\vspace{-1.81cm}

\unitlength1cm \epsfxsize=8.6cm   \epsfbox{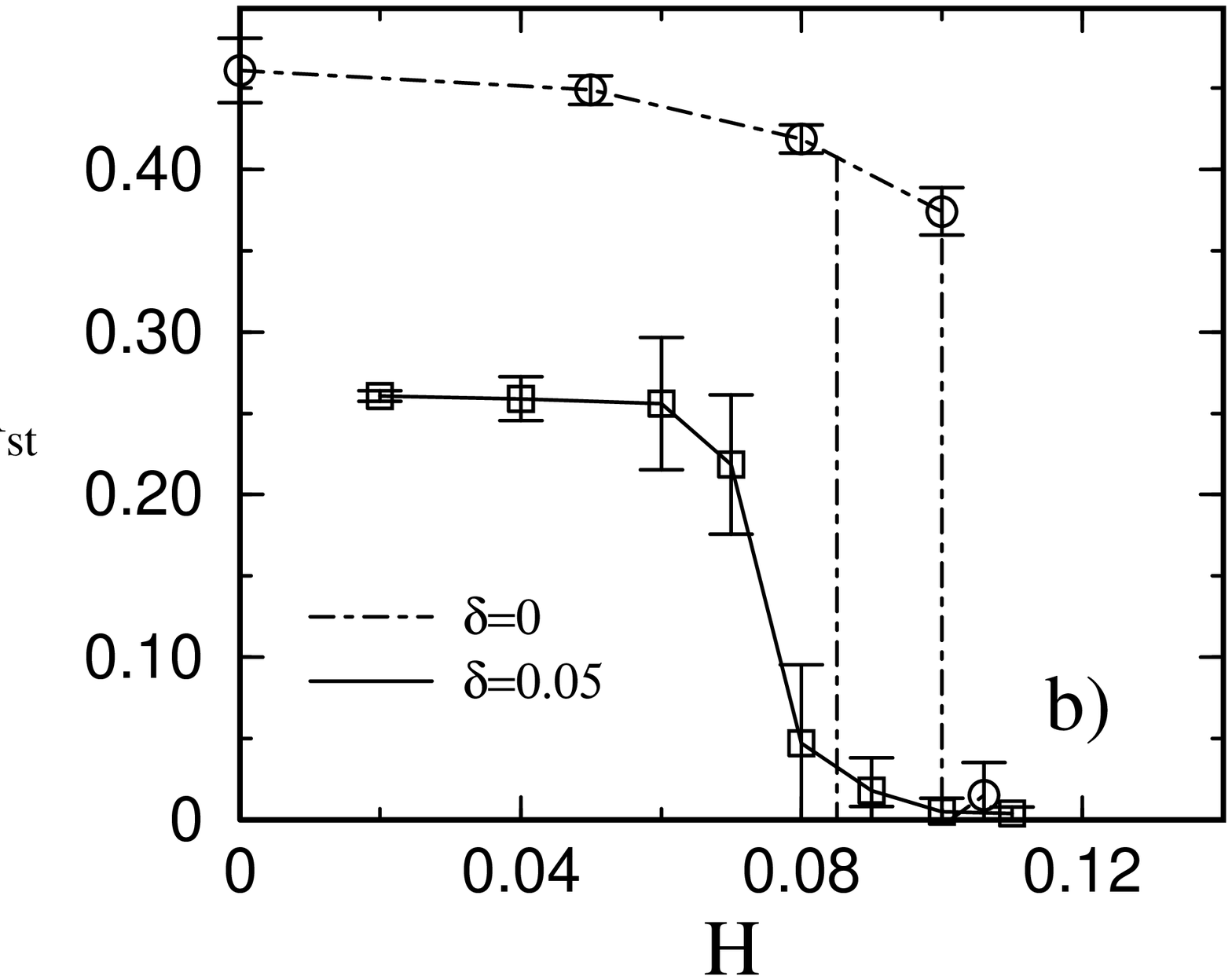}
\caption{Change of a) the magnetization $m(H)$, and b) the staggered 
magnetization $m_{st}(H)$ with doping for $U=2$ at $T=1/16$.
\label{mvsHdop}\\}
\end{figure}

Upon doping
the magnetization curve changes considerably and hardly indicates the
existence of a metamagnetic phase transition (Fig.~\ref{mvsHdop}).
This is due to the fact that in the metallic phase
there is no longer a ``Slater gap'' at the Fermi energy;
therefore the homogeneous susceptibility is not as
strongly affected by the antiferromagnetic order as at half filling. 
The phase transition is, however, clearly seen in $m_{st}(H)$.

\begin{figure}[t]
\unitlength1cm \epsfxsize=8.6cm   \epsfbox{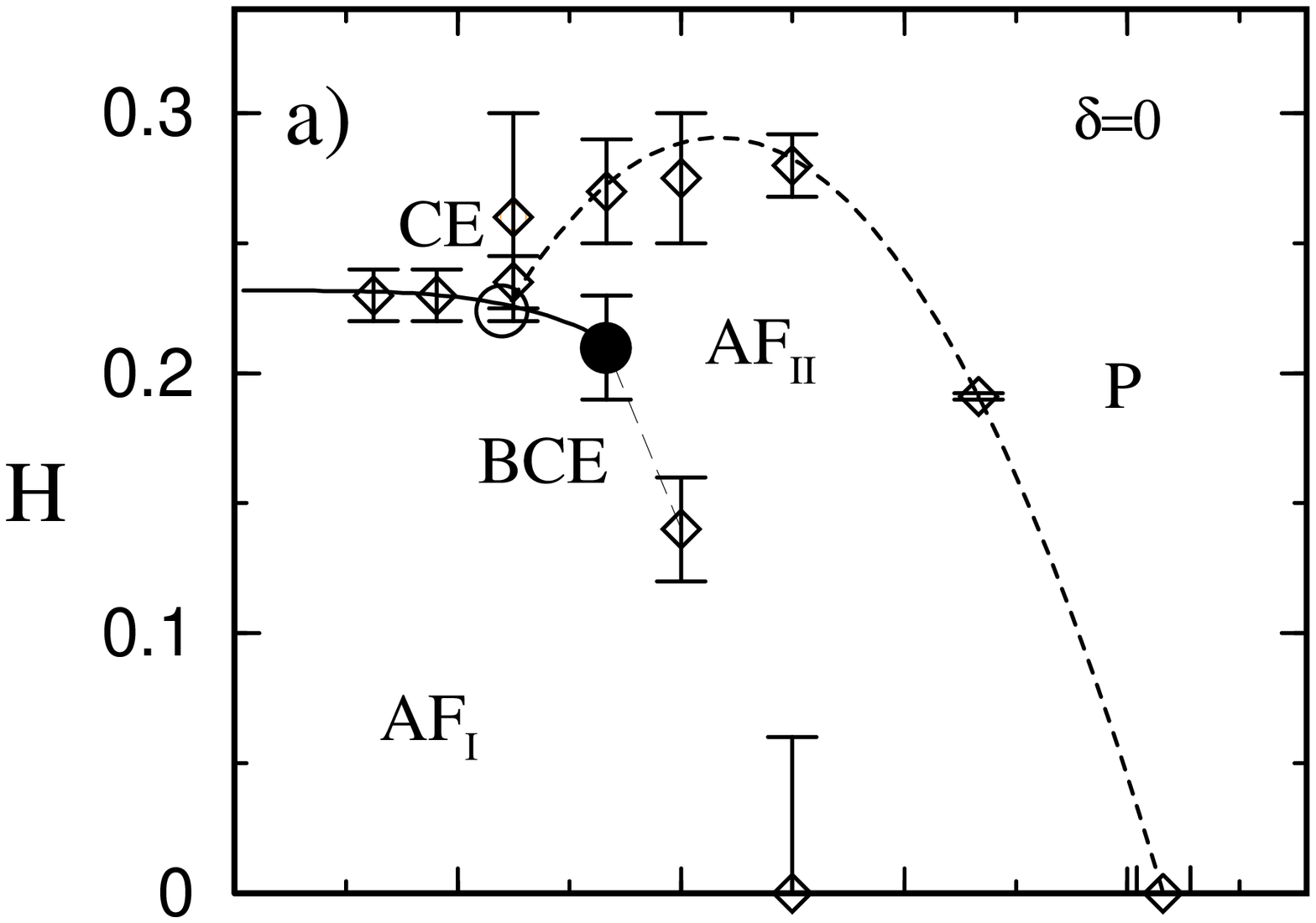}
\vspace{-1.81cm}

\unitlength1cm \epsfxsize=8.6cm   \epsfbox{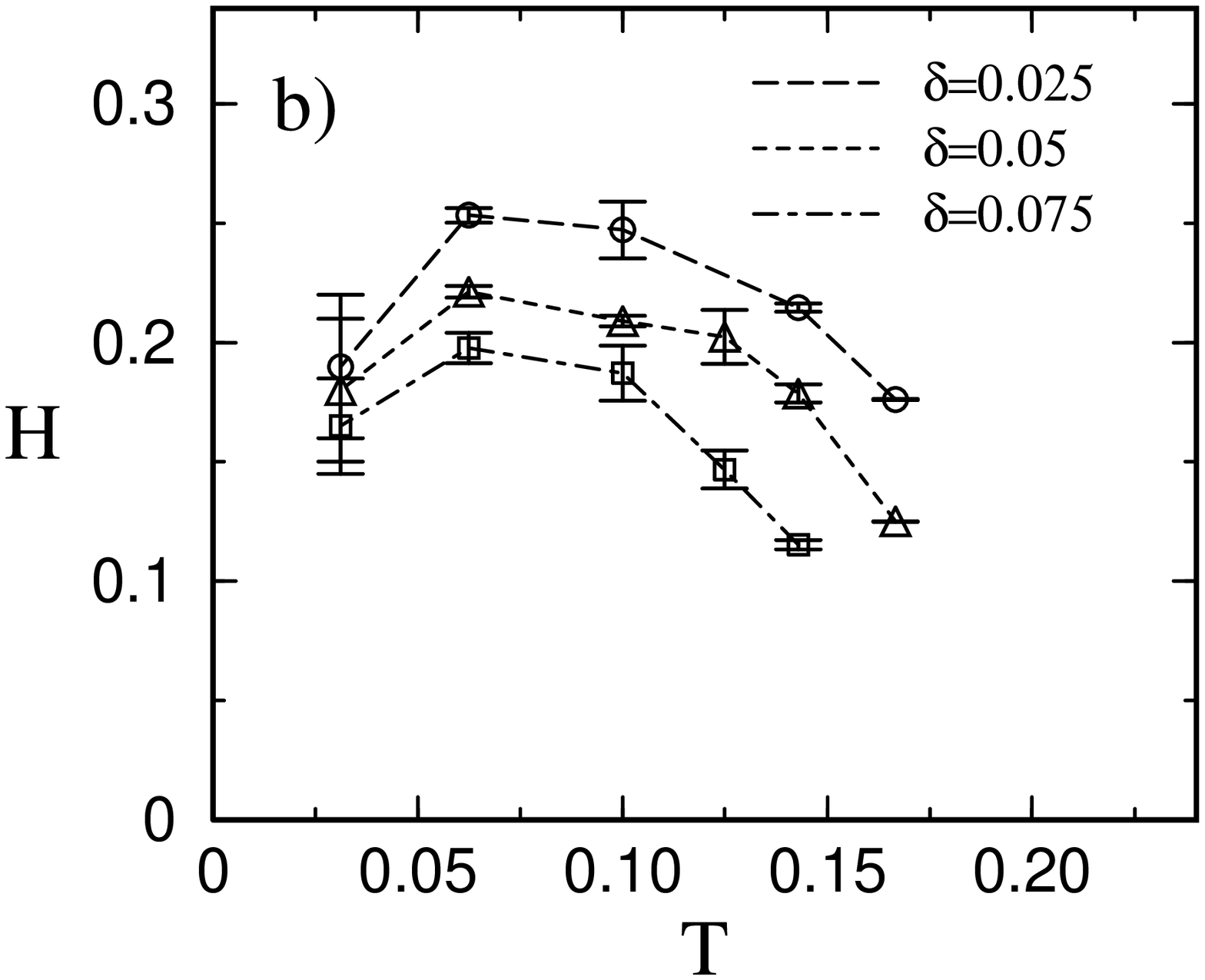}
\caption{Change of the $H-T$  phase diagram with doping $\delta$
for $U=4$. 
QMC data are shown for a) $\delta=0$,
b) $\delta = 0.025, 0.05,$ and $0.075$.
 Second order phase
transitions are indicated by  dashed lines, first
order transitions by solid lines.
For $T=1/32$ the numerical error does not permit the determination of
the order of the phase transition unambiguously.
\label{pddop}\\}
\end{figure}

From the $m_{st}$ vs. $H$ curve the phase diagram (Fig.~\ref{pddop}) is 
constructed. 
The metamagnetic phase transition line is found at lower temperatures
and fields compared to half filling. 

Associated with the metamagnetic phase transition is a change
of the electrical resistivity.
To study this important effect we calculated the $H$--dependence
of the electronic compressibility $\kappa_e$.
This quantity indicates whether the system is metallic or insulating.
For an insulator $\kappa_e$ vanishes for $T=0$ and
is exponentially small for temperatures lower than the antiferromagnetic gap.
In a Fermi liquid,  on the
other hand, $\kappa_e(T=0)$ is finite since it
is proportional to the density of states at the Fermi level and
hence proportional to the Drude conductivity.

\begin{figure}
\unitlength1cm
\epsfxsize=8.6cm 
 \epsfbox{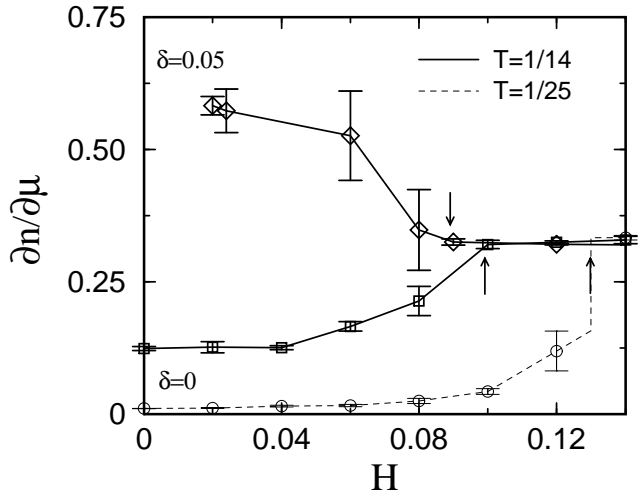}
\caption{Field dependence of the  electronic compressibility 
$\kappa_e = \partial n/\partial \mu$ 
for $U=2$ at $\delta = 0$ ($T=1/14,1/25$)
and $\delta=0.05$  ($T=1/14$). The arrows mark the respective
critical fields for the 
metamagnetic phase transition.
\label{comp}}
\end{figure}

\newpage
The results for $\kappa_e$ as a function of magnetic field $H$ at $U=2$
are shown in Fig.~\ref{comp}. At half filling, $\delta=0$, the compressibility
is seen to increase with $H$. This effect is
particular pronounced at low
temperatures $(T=1/25)$ where $\kappa_e$ is essentially zero at low
fields and rises to $\kappa_e\approx 0.3$ above the critical field,
indicated by an arrow. Hence the metamagnetic phase transition
is a transition from an antiferromagnetic insulator to a metal with
homogeneous magnetization.
At higher temperatures, $T=1/14$, the compressibility is always finite 
due to thermal excitations.  
We note that at $U=4$, when the electrons are essentially localized, 
$\kappa_e$ remains small 
($0<\kappa_e<0.03$ at $T=1/8$; not shown in Fig.~\ref{comp}) even above
the critical field, indicating an insulator--to--insulator transition.
The situation is very different at finite doping ($\delta=0.05$).
Here Fig.~\ref{comp} shows that $\kappa_e$ \em decreases \em with $H$
by approximately 50\% as the system goes through the metamagnetic transition
from an antiferromagnetic metal to a metal with homogeneous magnetization.

\section{Discussion}
\label{dis}
In summary,
we investigated the origin of metamagnetism in strongly
anisotropic antiferromagnets starting from 
a microscopic model of strongly
correlated electrons, the Hubbard model with easy axis,
by employing a dynamical mean--field theory.
This approach is {fundamentally} different from previous investigations 
since we identified and explicitly evaluated the simplest \em electronic, \em 
i.e.~fully quantum mechanical, correlation model that is able  to explain 
the conditions for metamagnetism.
For this electronic model we show unambiguously that at intermediate coupling
the phase transition 
is of first order at low temperatures and of second order near the N\'eel 
temperature, i.e.~the order of the phase transition {changes}.
This phenomenon, which has been of interest to various communities 
in classical 
statistical mechanics for a long time already, is extracted from a model of
itinerant electrons.

Our approach allows us to describe a broad range of  
qualitatively different metamagnets within a single model. 
While at present this simple model does not permit any quantitative
calculation of material properties it does describe 
itinerant and localized, metallic and insulating metamagnets and the crossover
between them. This crossover is related to two 
fundamental experimental parameters, i.e.~pressure 
(related to $U/t$ which decreases with pressure) and doping.

At half filling the  Coulomb interaction 
leads to a crossover from a band insulator to an insulator with
localized moments. Thereby the phase transition changes from
first order for the bandlike metamagnet to second order for the
localized one. Only at intermediate couplings are both first and second order
phase transitions observed as found in experiment.
The $H-T$ phase
diagram  obtained 
for an intermediate Coulomb interaction 
($U = 4$ = band width) is strikingly
similar to  that of FeBr$_2$\cite{Azevedo95,27} or the Ising model 
with $J^\prime \ll J$ \cite{8,Selke96}.
We note that in these insulating systems the applicability of a 
theory which becomes exact for large coordination number is 
justified by the fact that   the AF superexchange involves 20
equivalent sites in the two neighboring iron planes \cite{28}.
At smaller values of the Coulomb interaction ($U=2$)
the temperature regime with second order transitions
shrinks and the two step phase transition becomes less pronounced, 
reproducing the scenario of Fig.1a, as observed e.g~in FeCl$_2$ \cite{Str77}.

The calculations off half filling allow us to investigate the properties
of metallic metamagnets, such as the Uranium-based mixed-systems 
\cite{Str77,Sechovsky94}, for 
which a theory in terms of a correlated electron model is mandatory.
In contrast to the insulating case, the metamagnetic phase transition
in the metallic system is hardly visible in the magnetization curve. 
This is because there is no longer a gap at the Fermi energy.
Quite generally, the critical temperatures and fields
decrease upon doping.

The metamagnetic transition is accompanied by pronounced changes in
the conductivity of the system. The Hubbard model with easy axis can
qualitatively describe several scenarios:

(i) In the insulating, localized regime ($U\geq 4$ at half filling)
a magnetic field causes a transition from an antiferromagnetic insulator
to an insulator with homogeneous magnetization.

(ii) At lower $U$--values (e.g.~$U=2$) at half filling an insulator--to--metal
transition occurs at the magnetic field where the AF order disappears.
Such a phenomenon is observed, for example, in the AF phase of 
La$_{1-x}$Ca$_x$MnO$_3$, where the resistivity is found to change by
several orders of magnitude \cite{Schiffer95}.
This is referred to as ``colossal'' magnetoresistance. We note that 
La$_{1-x}$Ca$_x$MnO$_3$ shows no strong anisotropy. Therefore our approach
can only describe the general features, in particular the existence of the
insulator--to--metal transition.

(iii) Away from half filling a magnetic field induces a transition 
from a metallic antiferromagnet to a metal without staggered moment.
Here the compressibility changes by less than an order of magnitude,
e.g.~about 50\% at $U=2$, $\delta=0.05$.
A similar effect is found in several strongly anisotropic antiferromagnets,
both in multilayers and bulk intermetallic compounds such as UPdGe
\cite{Sechovsky94}. In these systems the origin of this 
``giant'' magnetoresistance is attributed to band structure effects
and spin scattering \cite{Levy95}. By contrast, our approach
stresses the importance of genuine electronic correlation effects.

More detailed investigations, including band degeneracy and spin--orbit
interaction, may eventually provide even quantitative insight into these
interesting and important phenomena.

\subsubsection*{Acknowledgments}
We acknowledge useful correspondence with N. Giordano, 
G. Lander, B. L\"{u}thi, A. Ramirez, and W. Wolf, and are grateful to 
H. Capellmann, P. van Dongen,  W. Metzner, 
H. M\"{u}ller-Krumbhaar, J.~Schlipf,
F.~Steglich, G. Stewart
and, in particular, to V. Dohm, W. Kleemann and W. Selke for very helpful
discussions. In its initial stages 
this work was supported in part by the SFB 341
of the DFG and by a grant from the ONR, N00014--93--1--0495.

\begin{appendix}

\section{Calculation of Susceptibilities from correlation functions}
\label{sus}
Quite generally susceptibilities can be obtained
from the derivative of the order parameter  $m_x$ w.r.t.~the
corresponding field $x$:
\begin{equation}
 \chi_x = \frac{\partial m_x}{\partial x} = 
 \frac{1}{2} T
 \sum_{\alpha  \sigma n} f_{\alpha  \, x}^{\sigma} 
\frac{\partial G_{\alpha n}^\sigma }{\partial x}
\label{susg}
\end{equation}

\begin{eqnarray}
\: \mbox{ with } & \:
 f_{\alpha \, x}^{\sigma} = 
\left \{ 
\begin{array}{c}
\sigma \\ \alpha \, \sigma \\ 1 \\ \alpha \\
\end{array}
\right\}
& \: \mbox{ for } \: x = 
\left \{
\begin{array}{c}
H \\ H_{st} \\ \mu  \\ \mu_{CDW} \\
\end{array}
\right\},
\end{eqnarray}
Here $x=H$ and $x=H_{st}$ 
lead to the ferromagnetic and antiferromagnetic 
susceptibilities,
$x=\mu$ to the  electronic compressibility,
and $x=\mu_{CDW}$ to the charge density wave susceptibility.
From the two self--consistency equations (\ref{dys}) and
(\ref{Gl4}) one obtains two corresponding equations for the 
derivative of the Green function w.r.t.~the variable $x$.
The derivative of the functional integral (\ref{Gl4}) gives:
\begin{equation}
\frac{\partial G^\sigma_{\alpha n}}{\partial x} = 
 T \sum_{\sigma' \, n'} \Gamma^{ \alpha
 \: \sigma \,\sigma'}_ {n n',n' n} \, \gamma^{\sigma' \: x}_{\alpha \, n'},
\label{gbar1}
\end{equation}
where $\Gamma$ is the local two particle correlation function
\begin{eqnarray} 
\Gamma^{ \alpha \: \sigma \,\sigma'}_ {n_1 n_1',n_2' n_2} &=&
< \! \Psi^\sigma_{\alpha n_1} {\Psi^{\sigma *}_{\alpha n_2}}
 \, \Psi^{\sigma '}_{\alpha n_1'} {\Psi^{\sigma ' *}_{\alpha n_2'}}
 \!> \\ 
 &&- \delta_{\sigma \sigma'}
< \! \Psi^\sigma_{\alpha n_1} {\Psi^{\sigma *}_{\alpha n_2}}
 \!> \;
<\! \Psi^{\sigma }_{\alpha n_1'} {\Psi^{\sigma *}_{\alpha n_2'}}
 \!>. \nonumber
\end{eqnarray}
The quantity
${\gamma^{ \sigma  \: x}_{\alpha n}} = \frac{\partial}{\partial x} 
\{ (G^\sigma_{\alpha n})^{-1}+\Sigma^\sigma_{\alpha n}\}$
 in (\ref{gbar1}) measures the response of
the averaged medium to an infinitesimal change of the
field $x$.
This dynamical response function
is determined by an integral equation in frequency space which does not
explicitly depend on momentum. (Note that there are no convolutions in
$\bf{k}$ space
in the $d = \infty$ limit as is typical for a mean--field
theory). This property does
not imply, however, that the response
function $\gamma_{\alpha n}^{ \sigma  \: x}$ is local, too. It
only indicates that $\gamma_{\alpha n}^{ \sigma  \: x}$ is {\em diagonal}
in the momentum $\bf{k}$. Momentum dependence enters implicitly
by the  particular $\bf{k}$ dependence of the external field
($\bf{k} = 0$
in the case of the compressibility or the ferromagnetic susceptibility,
and $\bf{k} = (\pi , \ldots , \pi)$ for the staggered susceptibility).

In the presence of an external field the variables $z^\sigma_{\alpha n}$ in
the  Dyson equation (\ref{dys}) are replaced by
$z^\sigma_{\alpha n}= i \omega_n+ \mu + \sigma H + \alpha \sigma H_{st} + 
\alpha \mu_{CDW}  - 
\Sigma^\sigma_{\alpha n}$. The derivative of the Green function yields:
\begin{eqnarray}
\frac{\partial G^\sigma_{\alpha n}}{\partial x}
  & = & \left \{ {\gamma^{\sigma \: x}_{\alpha n}} + 
\frac{{\partial G^\sigma_{\alpha n}}/{\partial x}}{(G^\sigma_{\alpha n})^2}-
f_{\alpha \, \,x}^\sigma \right \} \zeta^\sigma_{\alpha n}
\label{gs2}
+  \\ & &
\left \{ {\gamma_{-\alpha \, n}^{\sigma \: x}} + 
 \frac{{\partial G^\sigma_{\alpha n}}/{\partial x}}
      {{( G_{-\alpha \, n} ^\sigma )}^2}-f_{-\alpha \, \,x}^\sigma
\right \}  \eta^\sigma_{\alpha n}
 \nonumber 
\end{eqnarray}
with
\begin{eqnarray}
\zeta^\sigma_{\alpha n} &=& \int_{-\infty}^{\infty} \! d \epsilon \, D(\epsilon)
  \left( z_{\alpha n}^\sigma- \epsilon^2/z_{-\alpha
   n}^\sigma\right)^{-2},\\
 \eta^\sigma_{\alpha n}&=&\int_{-\infty}^{\infty} \! d \epsilon \, D(\epsilon)
   \left( z_{\alpha n}^\sigma - \epsilon^2/z_{-\alpha
   n}^\sigma\right)^{-2} \; \epsilon^2/(z_{-\alpha n}^\sigma)^2.
\end{eqnarray}
Since  (\ref{gs2}) separates in Matsubara frequencies $n$ and spin $\sigma$
it can be easily solved for $\partial G / \partial x$:

\begin{equation}
\frac{\partial G^\sigma_{\alpha n}}{\partial x} = \sum_{\alpha'} 
R_{ n}^{\sigma \; \alpha \alpha' } 
\left( f_{\alpha' \, \,x}^\sigma - \gamma^{\sigma \: x}_{\alpha' \, n} \right)
\label{gbar2}
\end{equation}
with the $2 \times 2$ array in  
$\alpha \in \{A,B\}$:
\begin{equation}
 {\mathbf R}_{ n}^{\sigma } = - ( {\mathrm det } \: \mathbf
 D)^{-1} {\mathbf D}^\sigma
_{ \, n} {\mathbf T}^\sigma_{ \, n},
\end{equation}
whereby $\bf D$ and $\bf T$ are defined as
\begin{eqnarray}
{\mathbf D}_{ n}^\sigma& =& \left( 
\begin{array}{cc}
1 - \frac{\zeta^{\sigma}_{B \, n}} {(G_{B \, n}^{\sigma})^2} &
\frac{\eta^{\sigma}_{A \, n}} {(G_{B \, n}^{\sigma})^2}  \\[8pt]
\frac{\eta^{\sigma}_{B \, n}} {(G_{A \, n}^{\sigma})^2}  &
1 - \frac{\zeta^{\sigma}_{A \, n}} {(G_{A \, n}^{\sigma})^2} \\
\end{array}
\right),\\[8pt]
{\mathbf T}^\sigma _{ \, n}&=&
\left( 
\begin{array}{cc}
\zeta^{\sigma}_{A \, n} &
\eta^{\sigma}_{A \, n} \\[8pt]
\eta^{\sigma}_{B \, n} &
\zeta^{\sigma}_{B \, n} \\
\end{array}
\right).
\end{eqnarray}

Now $\partial G/ \partial x$ can be eliminated by setting 
(\ref{gbar1}) equal to (\ref{gbar2}), yielding
\begin{eqnarray}
\lefteqn{ \sum_{\alpha'} R^{\sigma \: \alpha \, \alpha'}_{n} 
f_{\alpha' \, \,x}^\sigma \; =} \nonumber\\
& &\quad \sum_{\sigma' \alpha' n'} \left \{ \delta _{n n'} \delta_{\sigma
   \sigma'} R^{\sigma \: \alpha \, \alpha'}_{n} + \delta_{\alpha
   \alpha'} T  \Gamma^{ \alpha
 \: \sigma \,\sigma'}_ {n n',n' n} \right \} 
\gamma^{\sigma' \: x}_{\alpha' \, n'}.
\end{eqnarray}
From this equation we determine $\gamma$ by numerical inversion
of a $4 \Lambda \times 4 \Lambda$ matrix.
Knowing $\gamma$ we obtain $\partial G/ \partial x$
via (\ref{gbar1}) or (\ref{gbar2}) and thus
the susceptibility (\ref{susg}).

\newcommand{\mstperpm}{${\mathbf m_{st} \perp m}$}
\section{Hartree--Fock theory for \mstperpm}
\label{awc}
\label{perpo}
\label{hfo}
Similar to the derivation of the  Hartree--Fock equations for  
${\mathbf m} \parallel {\mathbf H} \parallel {{\mathbf m}_{st}}$ 
(see Sec.~\ref{wc})
we will now investigate
the case with perpendicular orientation 
${\mathbf m} \parallel {\mathbf H} \perp {{\mathbf m}_{st}}$.
The Ansatz for the one particle densities 
\begin{eqnarray}
  <\!n_{i \in \alpha \sigma}\!> = \frac {1}{2} ( n + 
  \sigma \; m) ,\vspace{3cm} <\!c_{i \in \alpha \sigma}^\dagger 
   c_{i \in \alpha \bar{\sigma}}^{\phantom{\dagger}}
  \! >  = \frac{1}{2}{\alpha m_{st}}
\end{eqnarray}
yields in addition to the Hartree term 
a Fock term in the decoupling (\ref{hde}) 
\begin{eqnarray}
    n_{i \sigma} n_{i \bar{\sigma}} &\stackrel{\mathrm HF}
   {\longrightarrow}&
 n_{i \sigma} <\! n_{i \bar{\sigma}}\! > +  <\! n_{i \sigma}\! > n_{i
   \bar{\sigma}}
 -  <\!n_{i \sigma}\!> <\!n_{i \bar{\sigma}}\!> \nonumber\\
 & & -  c_{i \sigma}^\dagger c_{i \bar{\sigma}}^{\phantom{\dagger}} 
 <\! c_{i \bar{\sigma}}^\dagger c_{i \sigma}^{\phantom{\dagger}} \! >
 - <\!c_{i \sigma}^\dagger c_{i \bar{\sigma}}^{\phantom{\dagger}} \! >
  c_{i \bar{\sigma}}^\dagger c_{i \sigma}^{\phantom{\dagger}} \nonumber\\
 && + <\! c_{i \sigma}^\dagger c_{i \bar{\sigma}}^{\phantom{\dagger}} \!>
 <\! c_{i \bar{\sigma}}^\dagger c_{i \sigma}^{\phantom{\dagger}} \! > . 
\end{eqnarray}
With this Ansatz one readily obtains the effective one--particle Hamiltonian
\begin{eqnarray}
 {\cal H}_{\mathrm {HF}} & &  = 
  \sum_{N N \sigma} t_{i j} \hat{c}_{i \sigma}^\dagger \hat{c}_{j
   \sigma}   - \frac{1}{2} \frac{U}{4} \sum_{i  \sigma} ( n^2 - m^2 
  -m_{st}^2) +  \\ && 
  \sum_{\alpha  i \in \alpha  \sigma} \!  \frac {U}{2} (n- \sigma
  m)  \hat{n}_{i \sigma}  -  \frac {U}{2}  \alpha m_{st} c^\dagger_{i \sigma} 
\!c^{\phantom{\dagger}}_{i \bar{\sigma}} - (\mu + \sigma \!H)
  \hat{n}_{i \sigma}.  \nonumber 
\end{eqnarray}
Diagonalizing this Hamiltonian yields the one--particle energies
\begin{equation}
\tilde{\epsilon}_{\sigma} \! = {\mathrm  sgn} \!\left({\scriptstyle \epsilon - 
\sigma \left\{\!\frac{U}{2} m+H\!\right\}} \right)
   \sqrt{\left({\scriptstyle \frac{U}{2}m_{st}}\right)^2+
   \left({\scriptstyle \epsilon - \sigma \left\{\!\frac{U}{2} m+H\! \right\}} 
\right)^2}.
\end{equation}
From these energies the free energy $\Omega$ is  calculated, and 
the minimization with respect to $m$ and $m_{st}$ leads to the 
following Hartree--Fock self--consistency equations
\begin{equation}
  m_{st} =  \frac{U}{2}\sum_\sigma \int {\mathrm d} \epsilon N^0(\epsilon)
  \frac{-  m_{st}}{ \tilde{\epsilon}_\sigma}  \frac{1}{1+
  e^{\beta (\tilde{\epsilon}_\sigma-\tilde{\mu})}} ,
\end{equation}
\begin{equation}
  m =   \sum_\sigma \int {\mathrm d}  \epsilon  N^0(\epsilon)
 \;\sigma \;
     \frac{\epsilon - \sigma \left(\frac{U}{2} m+H\right)}
{ { \tilde{\epsilon}_\sigma}  [1+
  e^{\beta (\tilde{\epsilon}_\sigma-\tilde{\mu})}]}.
\end{equation}
As in Sec.~\ref{wc} these Hartree--Fock equations are solved numerically.

\section{Series expansion of the Hartree--Fock free energy}
\label{icmhf}
The itinerant 
electron metamagnetism theory of Mori\-ya and Usami \cite{Moriya77} 
can be derived from the Hartree--Fock approximation only if
the free energy is analytic in $m$ and $m_{st}$. Since  in RPA the Hubbard 
interaction $U$  contributes to the free energy analytically 
(see (\ref{iemlandau}) and (\ref{RPA})) 
any non--analytic behavior must be due to the kinetic energy. Its
expansion in the order parameter $m_{st}$ at $T=0$ 
is analyzed in this section.

To calculate the expansion in $m_{st}$
a staggered magnetic field $H_{st}$ is introduced:
\begin{equation}
{\cal{H}} = \sum_{N N, \sigma} t_{i j} 
c_{i \sigma}^\dagger c_{j \sigma}^{\phantom{\dagger}}
- H_{st} \sum_{ \alpha, i\in\alpha, \sigma} \alpha \sigma n_{i \sigma}.
\label{HHST}
\end{equation}
On $A$-$B$ lattices 
the one--particle energies for this Hamiltonian show a gap at $\epsilon =0$
with square root singularities at its edge
\begin{equation}
\tilde{\epsilon} = {\mathrm sgn}( \epsilon) 
\sqrt {\epsilon^2 + H_{st}^2}.
\end{equation}

We consider the  half filled band, where 
the staggered magnetization $m_{st}$ is
calculated from the one--particle energies. It
shows the following asymptotic non--analytic
behavior for $H_{st} \rightarrow 0$:
\begin{eqnarray}
m_{st} & = & \sum_\sigma \int_{-1}^0  {\mathrm d}\epsilon N^0(\epsilon) 
\frac{-H_{st}}{\tilde{\epsilon}}\\
&=& 2 N^0(0) H_{st} \ln \left(1/H_{st}\right) + {\cal{O}}(H_{st}).
\label{eqnmst}
\end{eqnarray}
Similarly the asymptotic behavior of the  
energy   (\ref{HHST}) is obtained as
\begin{eqnarray}
\Delta E(m_{st}) &=& \sum_\sigma \int_{-1}^0 {\mathrm d}\epsilon 
N^0(\epsilon)( \tilde{\epsilon} -\epsilon) \\
&=& -\frac{1}{2} H_{st} m_{st} +  {\cal{O}}(H_{st}^2). 
\end{eqnarray}
Subtracting the contribution due to $H_{st}$,
$\Delta E_{\mathrm H_{st}}=-H_{st} m_{st}$, the asymptotic
dependence of the kinetic energy on $m_{st}$ reads
\begin{equation}
\Delta E_{\mathrm kin}(m_{st}) \stackrel{H_{st} \rightarrow 0}
{\longrightarrow}  \frac{1}{2}  \frac{m_{st}^2}{\ln (1/m_{st})}.
\end{equation}
This shows that $\Delta E_{\mathrm kin
}(m_{st})$
is non--analytic in $m_{st}$.
Therefore the itinerant electron
metamagnetism  theory cannot be derived from
the Hartree--Fock theory for the Hubbard model
 with an easy axis.

\section{The metamagnetic phase transition at strong coupling}
\label{asc}
In the limit of strong coupling and half filling the ${\cal{O}}(t^4/U^3)$
perturbation theory yields the effective spin Hamiltonian (\ref{Hsc}). 
In the following, we study the metamagnetic phase transition, and 
especially the order of 
the transition, for this effective Hamiltonian. Restricting ourselves 
to solutions with mixtures of ferromagnetic ($m$) and antiferromagnetic 
($m_{st}$) order,
the ground state energy is a polynomial in $m$ and $m_{st}$ ($t^*\equiv 1$):
\begin{eqnarray}
E & = & \frac {1}{2} \frac{1}{U} m^2
- { \frac {1}{2} \frac{1}{U}} m_{st}^2 + 
 \frac {1}{32} \frac{1}{U^3}
\left\{ 16\,{B}\,{m}^{2} - 16\,{B}\,m_{st}^{2} \right.  \nonumber \\
& & + {A}\,{m}^{4} - 2\,{A}\,{m}^{2}\,m_{st}^{2}
   - 2\,{A}\,{m}^{2} + {A}\,m_{st}^{4} + 6\,{A}\,
m_{st}^{2} \nonumber \\ 
& &\left. - 16\,{C}\,{m}^{2} - 16\,{C}\,m_{st}^{2}\right\}
 - {H} \,{m} + \mbox{const.}
\end{eqnarray}
One can see that the 
ferromagnetic next nearest neighbor interaction $C$ favors both 
saturated antiferromagnetism and saturated ferromagnetism 
rather than ferrimagnetic phases. By contrast the
plaquette term $A$ has contributions that
support the formation of a ferrimagnetic state.
To obtain the ground state the energy must be minimized with respect to $m$ and 
$m_{st}$ under the constraints $|m| \leq 1$ and $|m_{st}| \leq 1- |m|$.
Differentiation of $E$ with respect to $m_{st}$ shows that,  for fixed $m$,
$E$ has one maximum at $m_{st}=0$ and two minima at
\begin{equation} 
m_{st} = \pm {\sqrt \frac{8\,{U}^{2} + {
A}\,{m}^{2} - 3\,{A}   + 8\,{B} + 8\,{C}}{{A}}} .
\end{equation}
For sufficiently strong coupling $U$ (e.g. $U > 2$ in the case of   the 
hypercubic lattice with $A$=20, $C$=2 and $B$=4) these minima  are outside the 
constraint
$|m_{st}| \leq 1 - |m|$. Therefore $E$ becomes minimal at the border of the 
constraint, i.e.~for $|m_{st}| = 1 - |m|$. Replacing $m_{st}$ by $1-m$ 
the minimization with respect to $m$ readily yields for the ground state 
\begin{eqnarray}
m_{\phantom {st}} & = & \left\{ \begin{array}{l}
    0 \hspace{4em} \mbox{for} \hspace{1em}  H \leq \frac{1}{2} \frac{-A + 2 
\,C +2\,U^2+2\,B}{{ U}^3}\\[4pt]
1  \hspace{4em} \mbox{for}  \hspace{1em} H \geq \frac { - {C} + {U}^{2} + {B}}
{{U}^{3}} \\[4pt]
 \frac {{A} - 2\,{C} - 2\,{U}^{2} - 2\,{B} + 2
\,{H}\,{U}^{3}}{  {A} - 4\,{C}}  \hspace{2em}  {\mathrm else}\\
\end{array}
\right. \\
  m_{st} &=& 1 - m.
\end{eqnarray}

This ground state solution for the effective spin Hamiltonian shows 
a second order metamagnetic phase transition for $A>4C$. This is the case
for the 
hypercubic lattice ($A=20,C=2$) and for the Bethe lattice,
where  $A$=0 but $C=-1$, i.e.~the next nearest neighbor coupling is 
antiferromagnetic.
In conclusion, the strong coupling theory shows 
second order phase transition for all temperatures, even at $T=0$. 
\end{appendix}

\newpage

\newpage
\newpage
\newpage
\newpage
\newpage
\newpage
\newpage
\newpage


\begin{thebibliography}{99}
\bibitem{27} Note that the mixed--phase region in the phase diagram
of Ref.~[21] disappears when instead of the applied field $H_a$ the
internal field $H$ is plotted vs.~$T$.
\end{thebibliography}
\end{document}